\documentclass[lettersize,journal]{IEEEtran}
\IEEEoverridecommandlockouts
\usepackage{amsmath,amsfonts}
\usepackage{algorithmic}
\usepackage{algorithm}
\usepackage{array}
\usepackage{textcomp}
\usepackage{stfloats}
\usepackage{url}
\usepackage{verbatim}
\usepackage{graphicx}
\usepackage{cite}
\usepackage{adjustbox}
\usepackage[caption=false,font=footnotesize]{subfig}
\usepackage{tabularx}

\begin{document}

\title{Pre-Fault Voltage Discrimination and Time-Domain Protection for Distribution Networks with Inverter-Based Resources}

\author{Junyuan Zhao, Fran\c{c}ois~Bouffard,~\IEEEmembership{Senior Member,~IEEE}, and Géza~Joós,~\IEEEmembership{Life Fellow,~IEEE}
\thanks{This work was supported in part by InnovÉÉ, Montreal, QC, and the Natural Sciences and Engineering Research Council of Canada, Ottawa, ON.}
\thanks{J. Zhao and F. Bouffard are with the Department of Electrical and Computer Engineering, McGill University, Montreal, QC H3A~0E9, Canada and with the Groupe d'\'{e}tudes et de recherche en analyse des d\'{e}cisions (GERAD), Montreal, QC  H3T~1J4, Canada (emails: junyuan.zhao@mail.mcgill.ca; francois.bouffard@mcgill.ca).}
\thanks{G. Joós is with the Department of Electrical and Computer Engineering, McGill University, Montreal, QC H3A~0E9, Canada (email: geza.joos@mcgill.ca)}
}%

\markboth{IEEE TRANSACTIONS ON POWER DELIVERY}%
{Zhao \MakeLowercase{\textit{et al.}}: Pre-Fault Voltage Discrimination and Time-Domain Protection for Distribution Networks with Inverter-Based Resources} 


\maketitle 

\begin{abstract}
The increasing proliferation of inverter-based resources (IBRs) in distribution networks is presenting a major challenge for phasor-based overcurrent protection. This challenge stems from IBRs' lack of short-circuit current sourcing capacity. As a result, traditional overcurrent protection functions (e.g., ANSI 51) are inadequate in such scenarios, and warrant alternative approaches. Time-domain protection, for example, shows promise in overcoming this challenge. In this paper we propose a pre-fault voltage discrimination (PVD) strategy whose role is to detect faults and discriminate normal switching and transformer inrush disturbances from actual faults. The use of PVD allows for the design of a simple, yet effective fault detection algorithm by using time-domain protection principles for distribution networks containing IBRs. The introduction of PVD provides for faster fault detection without reducing security and dependability. Offline simulation experiments and controller hardware-in-the-loop real-time simulation validate the effectiveness of the proposed algorithm against various fault and normal switching events. 
\end{abstract}


\section{Introduction}
\IEEEPARstart{A}{ccurate} and fast fault detection is essential for reliable power system protection. Today's commercial relays are predominantly frequency-domain relays, that are phasor-based, with detection delays typically of one to one and a half cycles to allow for accurate signal estimation \cite{schweitzer2015speed}. Measurements based on the fundamental frequency component require a duration of at least one cycle for signals to settle down. Some phasor-based relays use a shorter data window of half a cycle at the cost of accuracy being traded off for speed \cite{schweitzer2015speed}. For faster accurate protection, transient components must be utilized. 

More importantly, relays responding to transients are inherently less affected by inverter-based resources (IBRs) \cite{schweitzer2015speed, muenz2024protection}. IBRs challenge traditional protection principles designed for networks with synchronous machines by more harmonic contents \cite{prabakar2021use, schweitzer2015speed} and lower short-circuit currents during faults \cite{prabakar2021use}, especially in systems dominated by IBRs running in the grid-forming (GFM) mode.

So-called \emph{time-domain protections}, which use actual voltage and current time-domain waveforms, have been developed and deployed chiefly for transmission system fault location \cite{schweitzer2015speed}. Time-domain protection products for distribution system applications are yet to emerge. However, we argue in this paper that there are applications like traveling waves of medium-voltage networks with deep IBR penetration.

\subsection{Literature Review}
\label{sc:lr}

In the context of fault detection in medium-voltage distribution networks with increasing proliferation of IBRs, three conditions are required to achieve \emph{dependability}—the ability of a relay to issue correctly trip commands for all faults it is expected to detect—, and \emph{security}—the ability of a relay not to issue trip commands when there are no faults. The first condition requires that the processing of signals should be able to extract precisely traveling waves (TWs) generated by the inception of a fault from the fundamental and harmonic components of the voltage and/or current signals being sampled at high frequencies \cite{prabakar2021use}.

Secondly, time-domain relays in medium-voltage distribution have to have settings that are far off from what is typically found in time-domain relays used in transmission. This is so because typical distribution lines are shorter than transmission lines, they have different $X/R$ ratios, and they are hosting significantly more IBRs than in transmission \cite{prabakar2021use}. The latter issue is the main motivation for this paper; our thesis here is that time-domain protection is both secure and dependable for fault detection in IBR rich systems because it is not affected by the low IBR fault currents.

Thirdly, it is critical for relays to be able to distinguish between fault inception TWs, and TWs produced by, for example, switching events and other normal network transients which include transformer inrush.

Tracing back to the 1970s, time-domain protection was proposed in \cite{dommel1978high} leveraging the fundamental physical phenomena associated with TW propagation. In the 1980s, a correlation-based technique was used for TW signal processing \cite{crossley1983distance,shehab1988travelling}, while in the 1990s the wavelet transform was used to extract TWs \cite{magnago1998fault}. However, those lacked the essential interpretation related to power system parameters \cite{prabakar2021use} and had no industrial use cases. With the advent of powerful machine learning algorithms, there have also been attempts using data-driven methods to detect and locate faults \cite{mansourlakouraj2021application, tashakkori2019fault, vaish2021machine}.

An industrial application of time-domain protection launched in 2016 with the first commercialization of a time-domain protection relay \cite{SELmanual} designed for transmission system fault detection and location. Among the protection elements used, traveling wave (TW) and incremental quantities were categorized as time-domain protection. The approach used in this commercial relay \cite{SELmanual} extracts TWs with a signal processing method called \emph{differentiator-smoother} \cite{schweitzer2015speed, schweitzer2016new}, within which the TW elements have to compromise between dependability and security when the fault initiating voltage is low. The incremental quantity elements in this relay, with the theoretical foundation established in the 1980s \cite{vitins1981fundamental}, operate slower than TW elements, but they can provide extra dependability and security. However, line impedance settings are required for incremental quantities, which can be hard to obtain accurately in distribution systems.

As mentioned in the third specification, a major obstacle with extending time-domain protection from transmission systems into distribution systems is the challenge to security associated with normal switching operations. Only a few research papers have discussed this issue. Among them \cite{michalik2006high, sarwagya2018high, lin2019discrete} propose specific elements without much of theoretical backing. Moreover, the issue related to the magnitude of transformer inrush currents, which can be several times those of normal operating currents \cite{greenwood1991electrical}, has rarely been discussed to this day. The commercial relay described in \cite{SELmanual} applies overcurrent supervision address the impact of switching operations. However, it could fail in distribution systems considering how high possible transformer inrush currents could be in comparison to normal operating currents.

\subsection{Contributions}

In light of the gaps between TW theory and its application in distribution system fault detection, a pre-fault voltage discrimination (PVD) strategy is proposed in this paper. It comprehensively utilizes fault transients in various time spans, applying different circuit models and approximations accordingly. Consequently, a time-domain protection approach for distribution systems is developed following the PVD strategy with two new TW operating quantities. In this spirit, our prominent motivation is to make the best use of TWs to speed up detection without compromising security or dependability by unifying the established frameworks of fault analysis at different time spans to produce added value in light of their complementary nature.

We explicitly divide fault development into three stages. Compared to existing methods, we collectively consider fault development from its initiation to its first detection at a measurement point, and then their reflections until steady state is reached. In contrast to the common practice of detecting TWs by their amplitudes, we detect the first TW arrival by identifying discontinuities in the first derivatives of voltages with respect to time. As this fault detection approach is independent of fault TW magnitudes, it makes it desirable for IBR protection considering their low fault currents. Moreover, it allows for TW detection, and thus fault detection, even when the TW was initiated at a low pre-fault voltage, which are synonymous with low TW amplitudes. 

Even though fault detection does not rely on high TW amplitudes here, the amplitude information is later used for fast tripping of faults with high TW amplitude relative to pre-fault voltages. A nonlinearity check is designed for TWs with low pre-fault voltage to differentiate them from those created by transformer inrush. For those faults initiating almost exactly at zero voltage, overcurrent supervision is used as a backup protection scheme, although in practice, faults initiated at a voltage of exactly zero is essentially impossible \cite{kasztenny2023line}. Simple threshold settings are proposed based on transformer inrush events without requiring prior knowledge of network parameters. 

The remainder of this paper is organized as such: Section~\ref{S II} introduces the physical modeling and assumptions of time-domain protection. Next, Section~\ref{sc:IV} proposes the PVD strategy and the associated protection algorithm. Section~\ref{sc:V} provides both proof of concept and performance evaluation of our proposal using electromagnetic transients program (EMTP) software, and execution results by control hardware-in-the-loop (C-HIL) real-time simulation (RTS) are given in Section~\ref{sc:HIL}. Finally, we offer conclusions and an outlook for future work in Section~\ref{conclusion}.

\section{Stages of Fault Evolution}
\label{S II}

In this section, we explicitly divide the time evolution of fault events into three stages. This division provides a natural decomposition of fault phenomena and their respective modeling as seen later in Section \ref{sc:approximations}, and it can guide time parameter design found in Section \ref{sc:algorithm}.

\emph{Stage I -- First fault wavefront:}
\label{sc:stgI} Stage I lasts from the fault's initiation to the arrival of the fault's corresponding first wavefront at the protection relay location.

In Stage I, the fault-induced TWs are modeled as step functions, and the TW amplitude is the negative of the pre-fault voltage value at the fault location \cite{de2021mathematical}. In addition, distribution lines have to be modeled using distributed line parameters to capture the TW propagation phenomena adequately.

\emph{Stage II -- Reflected TWs combining into lumped parameter line transients:}
\label{sc:stage ii} TW reflections can happen at any discontinuity points in the faulted circuit such as at the end of a line, conjunctions of lines, or at the interface of a lumped impedance in series or in shunt \cite{kimbark1949electrical}.

As TWs propagate and reflect, the transients expressed in terms of TWs bear considerable resemblance to transients in a lumped parameter line model, except that the currents and voltages vary by steps of the TW propagation distance divided by the speed of propagation \cite{kimbark1949electrical}.

\emph{Stage III -- Transients and steady state represented by a lumped parameter line model}
\label{stageIII}

After several round trips involving multiple reflections, TWs combine into stationary waves \cite{schweitzer2015speed}, and the equivalent circuit can be analyzed with a lumped parameter line model in the time domain.

\section{Time Domain Line Protection Algorithm}
\label{sc:IV}

\subsection{Pre-Fault Voltage Discrimination Strategy}
\label{sc:pvd}

TW elements in the commercial relay \cite{SELmanual} use gross changes in signal amplitudes as a way to detect anomalies in transmission system settings. To that effect, there is a need to set an amplitude threshold whose value is essentially a tradeoff between security and dependability. The approach depends on sufficiently large TW amplitudes, and loses dependability when the wave magnitude is small. Therefore, such TW elements for detection are not fully dependable \cite{SELmanual}  even in transmission systems. This limitation is amplified in distribution systems, where lower voltage levels result in smaller TW amplitudes which can be are easily masked by attenuation.

To overcome this limitation, a Pre-Fault Voltage Discrimination (PVD) strategy is proposed in this paper. PVD is an adaptive time-domain strategy whose fault discrimination criterion is based on the ratio of TW amplitude to the pre-fault voltage (TWPV). This approach enables rapid detection for faults with high TWPV while applying additional verification for those with low TWPV. Instead of relying solely on amplitude, it first detects the presence of a transient using the rate of change of voltage ($dV/dt$), then evaluates the TW amplitude relative to the pre-fault voltage: 


\begin{itemize} 

\item  If the TWPV is high, the relay asserts immediately when the first TW wavefront is captured. 

\item  If it is low, an additional verification delay is applied before tripping. 

\item  When the pre-fault voltage is near zero, a longer delay is used as backup protection, still within one cycle. 

\end{itemize} 

Compared with the amplitude-based method \cite{SELmanual}, PVD separates detection (via rate-of-change) from discrimination (via TWPV), and it avoids compromising either security or sensitivity, using a longer time delay as a tradeoff in the case of low TWPV.

\subsection{Relevant Approximations}
\label{sc:approximations}
\subsubsection{Approximation of pre-fault dc source during Stage I}
\label{assumption dc}

As noted in Section \ref{sc:stgI}, the voltage source related to a fault is commonly modeled as a dc source. In this paper, the grid-side source during Stage I is likewise approximated by a dc source equal to its pre-fault value, instead of a sinusoidal source.

This simplification is justified because the sinusoidal voltage changes very little over the short time window relevant to TW propagation. For a propagation speed of $3 \times 10^8$ m/s, a $100~\mu$s interval corresponds to 30 km, longer than most distribution feeders, while a 60~Hz waveform changes by at most $3.77\%$ over the same duration.

The significance of this approximation is that the first voltage TW can be obtained directly by subtracting the pre-fault dc value, preserving its characteristics such as magnitude and rate of change, rather than an output of filters such as the differentiator-smoother filter used in commercial relays \cite{schweitzer2016new} or the discrete wavelet transform in some academic research \cite{michalik2006high, lin2019discrete, galvez2020fault} which come with lower interpretability and higher computational burden.

\subsubsection{Approximation of the pre-fault voltage at the fault location}
\label{sc:approximation}

Since the pre-fault voltage value at the exact fault location is unknown, this algorithm uses the voltage value at the measurement point as an approximation of that voltage. In a typical distribution network, like, for example, the IEEE 34 Node Test Feeder \cite{ieee34}, the load flow results show that the largest phase angle difference for phase $a$ between nodes, $\delta$, does not exceed $4^\circ$. In fact, the highest voltage error $\Delta V$ caused by such an approximation is
\begin{align}
        \max{\Delta V} &= \sqrt{2} V_M \max{\{ \sin (\omega t+\theta) - \sin (\omega t+\theta-\delta) \}} \nonumber \\ 
        &= \sqrt{2} V_M \max{\{2\sin{\delta/2}  \cdot \cos(\omega t+\theta-\delta/2)\}} \nonumber \\
        &=2 \sqrt{2} V_M \sin{\delta / 2} \label{eq:approximation}
\end{align}
For a voltage angle difference of $\delta = 4^\circ$, the corresponding voltage difference is $0.07\cdot\sqrt{2} V_M$. Although this difference is small, it will be explicitly taken into account when designing the detection algorithm in Section~\ref{sc:step ii}.


\subsection{Proposed Algorithm}
\label{sc:algorithm}

\begin{figure*}[!htbp]
    \centering
    \includegraphics[width=1\linewidth]{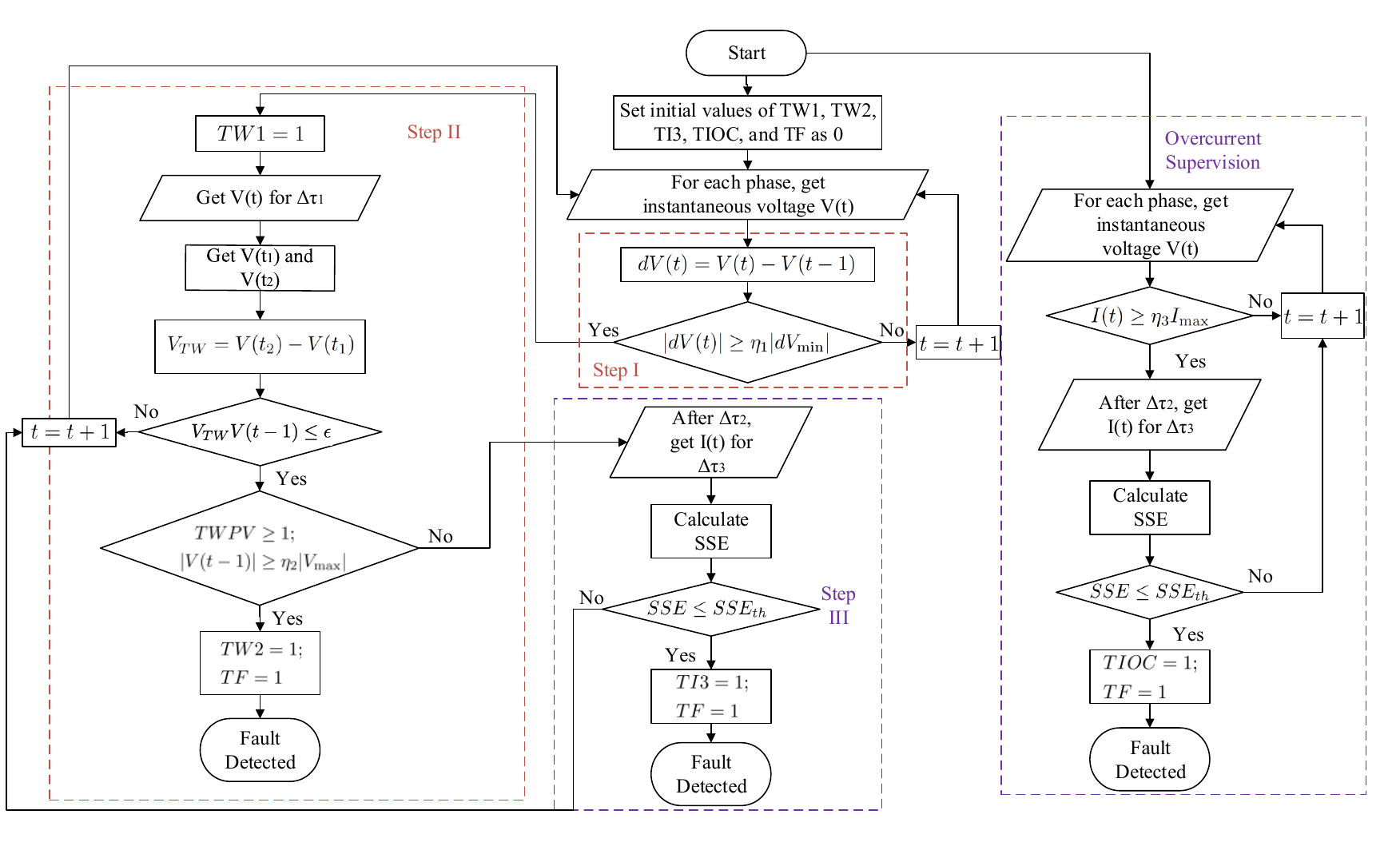}
    \caption{Flowchart of the fault detection algorithm.}
    \label{fig:flowchart}
\end{figure*}

The proposed algorithm is illustrated in Fig. \ref{fig:flowchart}, and triggering of each protection element, i.e. $TW1$ (TW detected), $TW2$ (TW not a switching event), $TI3$ (not transformer inrush), $TIOC$ (detection by overcurrent supervision) and $TF$ (fault detected), is explained in detail in the following subsections.

\subsubsection{Step I -- Traveling Wave Detection}
\label{sc:step1}

TWs can be generated by faults, and by other power system switching events where there is a change of path for the current. The first step of the algorithm is therefore to detect any TW at the relay terminals. Most TW detection methods \cite{michalik2006high,sarwagya2018high,lin2019discrete,schweitzer2016new,schweitzer2016performance} extract a TW by filtering and detecting it by its amplitude. However, faults initiating at a low voltage angle can generate TWs of low amplitude \cite{de2021mathematical, SELmanual}. This implies that a proper threshold setting is a compromise between security and dependability, potentially leading to considerable detection errors. 

On the other hand, the rate of change of voltage ($dV/dt$) in the presence of a fault TW can be higher than pre-fault values, regardless of the TW amplitude. In the proposed approach, the first operating quantity $dV$ is defined as the instantaneous voltage value less that of the previous time sample as in \eqref{eq:dvop}. The minimum operating threshold $|dV_{\min}|$ is defined as the maximum historical ($t \in H = [t_0, t_0 + 1, \ldots ,t-2,t-1]$) value of $|dV(t)|$ during normal steady-state operation. If $|dV(t)|$ is greater than or equal to $\eta_1 |dV_{\min}|$ \eqref{eq:dvcr}, where $\eta_1 \geq 1$ is user-defined, the traveling wave detection element $TW1$ is raised to 1, and Step II is triggered. 
\begin{align} 
    dV(t) &=  V(t)-V(t-1) 
    \label{eq:dvop}  \\
    |dV_{\min}| &=  \max_{t \in H}{|dV(t)|} 
    \label{eq:dvmin} \\
   |dV(t)| & \geq \eta_1 |dV_{\min}| 
    \label{eq:dvcr} 
\end{align}

\subsubsection{Step II -- Fast Tripping for Non-Switching Event}
\label{sc:step ii}

In Step I, TW arrival is detected using $dV$, which is more dependable than amplitude-based detection. Once a TW occurrence is confirmed, its amplitude $V_{TW}$ can be safely extracted, even if small, and compared with the pre-fault voltage value $V(t-1)$ to exclude normal switching events from faults. The TW magnitude is computed from the voltage signal within a short window of $\Delta \tau_1 = 100 \mu$s following the initial disturbance detection, as defined in \eqref{eq:VTW}. In \eqref{eq:VTW}, $V(t_1)$ and $V(t_2)$, with $(t_1, t_2) \in \Delta \tau_1$ such that $t_1 < t_2$, correspond to the samples leading to the largest $V_{TW}$ observable within the $\Delta \tau_1$ window.
\begin{equation} 
    V_{TW} =  V(t_2)-V(t_1) 
    \label{eq:VTW} 
\end{equation} 



Both distribution line faults and line switching events can cause sudden changes in voltage, thus generating TWs\cite{schweitzer2014locating}. The difference between switching and fault events resides in the amount of stored energy in the line capacitance prior to the disturbance. For switching events, the TW amplitudes are small relative to the pre-fault voltage, since they release far less energy than faults. Conditions \eqref{eq:operating1}--\eqref{eq:VTWfastmin} define the operating area of fault events in the possible presence of line switching. First, during faults, the polarity of the TW is opposite to that of the pre-fault voltage $V(t-1)$ \cite{de2021mathematical}, as shown in \eqref{eq:operating1} where $\epsilon < 0$ and $|\epsilon|\ll 1$. Second, we define $TWPV$ in \eqref{eq:VTWfast} as the ratio of the absolute value of the TW amplitude to the absolute value of the pre-fault voltage. In theory, fault TWs have $TWPV = 1$; however, in practice, attenuation makes $TWPV$ slightly below one, while measurements often show overshoot or ringing near the wavefront due to reflections and sensor dynamics \cite{kasztenny2023dependability}. Therefore, using a practical threshold of $TWPV \geq 1$ is appropriate here. Finally,  \eqref{eq:VTWfastmin} sets a constraint for low fault initiating angles around zero, to prevent false alarms that may occur when numerical errors become significant compared to extremely small signals and also to take into account errors brought by the approximations discussed in Section \ref{sc:approximation}. In \eqref{eq:VTWfastmin}, $|V_{\max}|$ is the maximum value of $|V(t)|$ observed in normal operations, and $\eta_2 \in (0,1)$ is another user-defined parameter.
\begin{align}  
    V_{TW} V(t-1) &\leq \epsilon 
    \label{eq:operating1} \\
    TWPV = \frac{|V_{TW}|}{|V(t-1)|} & \geq 1
    \label{eq:VTWfast}  \\  
    |V(t-1)|  &\geq \eta_2 |V_{\max}|
    \label{eq:VTWfastmin}  
\end{align} 
If all three conditions \eqref{eq:operating1}--\eqref{eq:VTWfastmin} are satisfied, $TW2$ is raised to 1 and, as a result, the on-fault trip signal, $TF = 1$, is generated.

\subsubsection{Step III -- Checking Cases Triggering TW1 but not TW2}
Step III is used to exclude transformer inrush from fault events for cases triggering $TW1$ but not $TW2$. Traditional differential protection schemes can block or restrain relay functions during transformer inrush by detecting high second harmonic contents associated with those currents. High second harmonic currents can be hard to detect and can cause misoperations. In addition, accurate harmonic analysis requires a much longer time window than time-domain protection. To remedy this shortfall, an alternative time-domain method is proposed for faster discrimination.

The current expected in Stage III is the response of an RL circuit to a sudden short-circuit
\begin{gather}
    i(t)  =\frac{\sqrt{2} V_M}{\sqrt{R^2+(\omega L)^2}} \left \{ \sin \left [ \omega t + \theta - \tan ^{-1} \left ( \frac{\omega L}{R} \right ) \right ] \right. \nonumber \\
     \qquad \qquad \qquad \qquad \left. -\sin \left [ \theta-\tan ^{-1} \left ( \frac{\omega L}{R} \right ) \right ] e^{-\frac{R}{L} t} \right \}
    \label{eq:I3}
\end{gather}
where $R$ and $L$ are the circuit resistance and inductance, respectively \cite{van2001transients}. We can rewrite \eqref{eq:I3} as \eqref{eq:AI3} to separate constant values and functions that vary over time
\begin{align}
    i(t) & =A\sin (\omega t+\alpha) - B e^{-\frac{R}{L} t} \nonumber \\
    & = A \cos \alpha \cdot \sin \omega t + A \sin \alpha \cdot \cos \omega t + B e^{-\frac{R}{L} t}
    \label{eq:AI3}
\end{align}
where
\begin{align}
    A &= \frac{\sqrt{2} V_M}{\sqrt{R^2+(\omega L)^2}}\\
    B &= -A \cos \alpha \label{eq:AI3param} \\
    \alpha &= \theta-\tan ^{-1} \left (\frac{\omega L}{R} \right )
\end{align}
For a time window $\Delta \tau_3$ within Stage III with samples at $t_1, t_2, \ldots, t_n$, current samples are gathered as
\begin{equation}
    \begin{bmatrix}
i(t_1) \\
i(t_2) \\
\vdots \\
i(t_n)  
\end{bmatrix} = \begin{bmatrix}
\sin \omega t_1 & \cos \omega t_1 & e^{-\frac{R}{L} t_1} \\
\sin \omega t_2 & \cos \omega t_2 & e^{-\frac{R}{L} t_2} &  \\
\vdots & \vdots & \vdots\\
\sin \omega t_n & \cos \omega t_n & e^{-\frac{R}{L} t_n} 
\end{bmatrix} \begin{bmatrix}
A\cos\alpha \\
A\sin\alpha \\
B \end{bmatrix}
\label{eq:matrixform}
\end{equation}
otherwise put
\begin{equation}
    \mathbf{i}(\mathbf{t})=\mathbf{\Theta(t)} \boldsymbol{\xi}
    \label{eq:AI3E}
\end{equation}
in which $\mathbf{i}(\mathbf{t})$ is the measured current data, $\mathbf{\Theta}(\mathbf{t})$ is the functional matrix, and $\boldsymbol{\xi}$ is the constant vector. The constant vector $\boldsymbol{\xi}$ can be estimated using a least squares fit
\begin{equation}
    \hat{\boldsymbol{\xi}} = (\mathbf{\Theta}^\top \mathbf{\Theta})^{-1}\mathbf{\Theta}^\top \mathbf{i}(\mathbf{t})
    \label{eq:ls}
\end{equation}
and the corresponding estimated currents are
\begin{equation}
    \hat{\mathbf{i}}(\mathbf{t})=\mathbf{\Theta} \hat{\boldsymbol{\xi}}
    \label{eq:ihat}
\end{equation}
Finally, the sum of squared errors (SSE) associated with the least-squares fitting is
\begin{equation}
    SSE = \sum_{k=1}^{n} (i(t_k)-\hat{i}(t_k))^2
    \label{eq:err}
\end{equation}

In contrast to a fault, the overcurrent generated by transformer inrush is due to the magnetic characteristic of the transformer core and corresponds to the nonlinear region in the flux density and magnetizing curve \cite{kasztenny2021traveling}. Therefore, when an inrush event is observed, the $SSE$ in \eqref{eq:err} is expected to be much higher than in the case of a fault event. If $SSE$ is less than some threshold $SSE_{th}$, the algorithm infers that the circuit response is indeed linear and, as a result, $TI3$ and $TF$ are both set to 1.

Step III is applied on a time window $\Delta \tau_3$ set to one third of a cycle of the fundamental. It should be noted that there is a $\Delta \tau _2 = 0.003$~s delay before the execution of Step III in Fig. \ref{fig:flowchart}. This is to reduce the influence of reflected TWs expected in Stage II. This delay corresponds to a traveling distance of 90~km for TWs moving at the speed of light, which is sufficient for a typical distribution network.



\subsubsection{Overcurrent Supervision}
\label{sc:oc} A time -domain overcurrent supervision is also added as a backup protection as shown in \eqref{eq:octh}. The protection function here is driven by the condition
\begin{equation}
    |I(t)| \geq \eta_3 I_{\max}
    \label{eq:octh}
\end{equation}
in which $\eta_3 \geq 1$ is a user-selected coefficient and $I_{\max}$ is the maximum current during normal operation. The value of $\eta_3$ can be set low, such as 2, since fault currents in distribution systems can be less significant than in transmission systems. With a low threshold setting, transformer inrush currents can also be captured. Therefore, the inrush discrimination criterion introduced earlier in \eqref{eq:err} is applied again in this step, and a delay $\Delta\tau_2$ is likewise added for security. In the event where these conditions are met, $TIOC$ and $TF$ are set to one.


\section{Performance Evaluation in Offline Simulation}
\label{sc:V}

In this section, we demonstrate the viability of the proposed algorithm on a distribution network with grid-connected IBRs using commercial software \cite{emtprv} with conservative parameter settings.

\subsection{Distribution Network under Test}
\begin{figure*}[!htbp]
    \centering
    \includegraphics[scale=0.14]{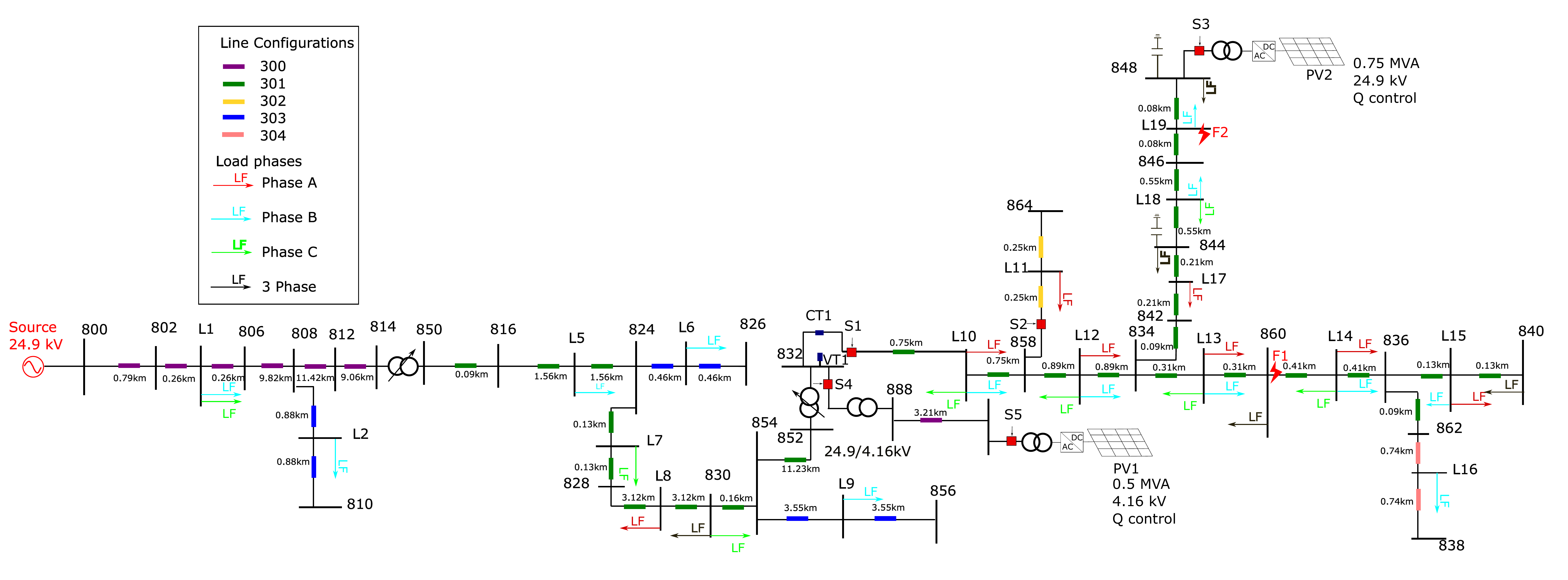}
    \caption{Modified IEEE 34 node test feeder.} 
    \label{fig:modifiedIEEE34_export} 
\end{figure*}

Tests are carried out on the modified IEEE 34 node test feeder shown in Fig.~\ref{fig:modifiedIEEE34_export}. The IEEE 34 Node Test Feeder is an actual 24.9~kV feeder located in Arizona \cite{ieee34}. It has two voltage regulators along the long distribution lines, and one 24.9/4.16~kV transformer connected to a short line section. The system is lightly loaded with unbalanced loading of both spot and distributed loads. Distributed loads are represented as loads connected at the center of line sections.

The modified model adds two inverter-based photovoltaic (PV) units \cite{galvez2020fault} connected at buses 848 and 890, providing 750~kW and 500~kW, respectively. The PV units work in voltage control mode. The transformers connected with the two PV units and the one at bus 888 are saturable models to test the robustness under transformer inrush during energization. The distribution line is represented by a constant distributed parameter model to enable TW observation \cite{borghetti2008continuous}. The sampling rate used is 1~MHz, a rate commonly used in commercial transmission time-domain relays \cite{SELmanual}.

\subsection{Simulation Cases and Results}
\label{sc:simulation}

Normal switching cases for line and transformer energization and de-energization are tabulated in Tables~\ref{tb:sw_feeder} and \ref{tb:sw_pv}. Faults with different fault resistance values $R_f$, fault inception angles and fault types are simulated and tabulated in Table~ \ref{tb:f}. We consider two fault locations: F1 (Bus 860) and F2 (Bus L19).

\begin{table*}
\centering
\caption{Normal Switching Cases: Feeder Energization and De-energization}
\begin{tabular}{c|cc|ccccc}
\hline
Event type        & \multicolumn{2}{c|}{Feeder de-energization} & \multicolumn{5}{c}{Feeder energization}                              \\ \hline
No.               & SW1                  & SW4                  & SW2        & SW3        & SW2\_1       & SW2\_2       & SW2\_3       \\
Switch            & S2                   & S4                   & S2         & S4         & S2           & S2           & S2           \\
Operation         & Open                 & Open                 & Close      & Close      & Close        & Close        & Close        \\
Load & 2~kW, 1~kvar           & 2~kW, 1~kvar           & 2~kW, 1~kvar & 2~kW, 1~kvar & 20~kW, 10~kvar & 20~kW, 10~kvar & 20~kW, 10~kvar \\
Switching angle ($^\circ$)   & 0                    & 0                    & 0          & 0          & 0           & 90            & $-90$          \\ \hline
\end{tabular}
\label{tb:sw_feeder}
\end{table*}

\begin{table*}
\centering
\caption{Normal Switching Cases: PV Energization and De-energization}
\begin{tabular}{c|ccccccc|cc}
\hline
Event type      & \multicolumn{7}{c|}{PV transformer energization}               & \multicolumn{2}{c}{PV transformer de-energization} \\ \hline
No.             & SW5     & SW7    & SW5\_1 & SW5\_2 & SW5\_3  & SW5\_4 & SW5\_5 & SW6                      & SW8                     \\
Switch          & S3      & S5     & S3     & S3     & S3      & S3     & S3     & S3                       & S5                      \\
Operation       & Close   & Close  & Close  & Close  & Close   & Close  & Close  & Open                     & Open                    \\
PV capacity     & 0.75~MVA & 0.5~MVA & 7.5~MVA & 75~MVA  & 0.75~MVA & 75~MVA  & 75~MVA  & 0.75~MVA                  & 0.5~MVA                  \\
Switching angle ($^\circ$) & 0       & 0      & 0      & 0      & 10      & 90     & $-90$    & 0                        & 0                       \\ \hline
\end{tabular}
\label{tb:sw_pv}
\end{table*}

\begin{table}[!htb]
\centering
\caption{Fault cases}
\begin{tabular}{llll}
\hline
No.    & Location & Type & $R_f$ and Fault Angle         \\ \hline
FT1    & F1       & AG   & $5~\Omega$ and $0^\circ$    \\
FT2    & F1       & AB   & $10~\Omega$ and $10^\circ$  \\
FT3    & F1       & ABG  & $15~\Omega$ and $45^\circ$  \\
FT4    & F2       & ABC  & $20~\Omega$ and $90^\circ$  \\
FT5    & F2       & ABCG & $20~\Omega$ and $135^\circ$ \\
FT6    & F2       & AG   & $20~\Omega$ and $10^\circ$  \\
FT6\_1  & F2       & AG   & $20~\Omega$ and $-15^\circ$ \\
FT6\_2  & F2       & AG   & $20~\Omega$ and $-58^\circ$ \\
FT2\_2 & F1       & AB   & $10~\Omega$ and $-58^\circ$ \\ \hline
\end{tabular}
\label{tb:f}
\end{table}

\begin{figure}
    \centering
    \subfloat[]{
        \includegraphics[
            width=0.47\linewidth,
            trim=3.5cm 8.5cm 3cm 9cm,
            clip
        ]{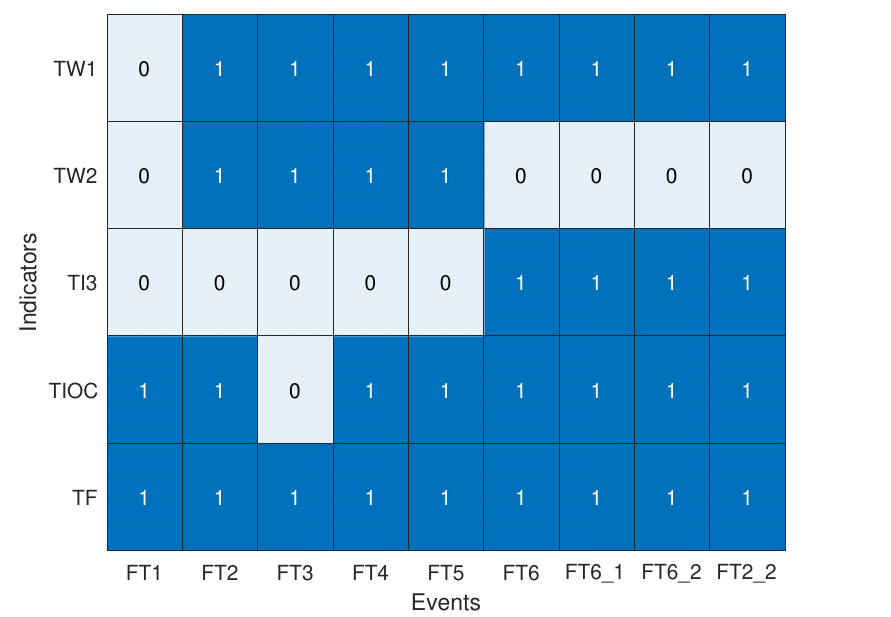}
        \label{fig:result_f}
    }
    \hfill
    \subfloat[]{
        \includegraphics[
            width=0.47\linewidth,
            trim=3.5cm 8.5cm 3cm 9cm,
            clip
        ]{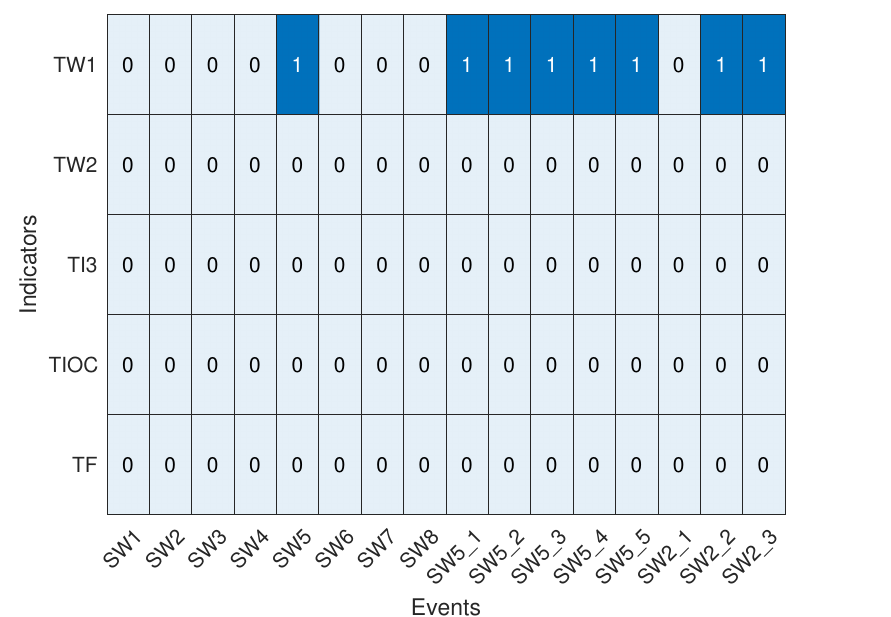}
        \label{fig:result_s}
    }
    \caption{Detection of (a) fault cases (b) switching events.}
    \label{fig:results}
\end{figure}

The results are collectively shown in Fig. \ref{fig:results}, including results for each protection element, i.e. $TW1$, $TW2$, $TI3$, $TIOC$ and $TF$. For the purpose of this demonstration, $\eta_1 = 5$, $\eta_2 = 0.10$, $\eta_3 = 2$, and $SSE_{th}$ is set to $80\%$ of the lowest SSE of all the tested PV transformer energization events, which is 38 in this network. These settings could be adjusted if required based on the system operator's preferences.

Fig. \ref{fig:results} shows how the proposed algorithm successfully identifies all cases correctly. The results are analyzed in detail further in Sections~\ref{sc:df}--\ref{sc:id}.

A summary of fault detection times is given in Table~\ref{tb:time}, where  $dT_f$ is the fault detection time, $dT_{TW2}$, $dT_{TI3}$ and $dT_{OC}$ are the detection times of $TW2$, $TI3$ and $TIOC$, respectively. Faults triggering $TW2$ can be detected in tens of microseconds, while faults which trigger $TI3$ can be detected in about half a cycle. Faults only triggering $TIOC$ can be detected within a cycle. Clearly, the utilization of TW information can greatly speed up fault detection.

\begin{table}[]
\centering
\caption{Fault Detection Times}
\begin{tabular}{ccccc}
\hline
Fault    & $dT_f$ (ms) & $dT_{TW2}$ (ms) & $dT_{TI3}$ (ms) & $dT_{OC}$ (ms) \\ \hline
FT1    & 10.009     & -               & -               & 10.009        \\
FT2    & 0.036      & 0.036          & -               & 9.099         \\
FT3    & 0.037      & 0.037          & -               & -              \\
FT4    & 0.039      & 0.039          & -               & 8.572         \\
FT5    & 0.040      & 0.040          & -               & 8.576         \\
FT6    & 8.571      & -               & 8.571          & 9.707         \\
FT6\_1 & 8.571      & -               & 8.570          & 10.825        \\
FT6\_2 & 8.571      & -               & 8.571          & 13.282        \\
FT2\_2 & 11.629     & -               & -               & 11.629        \\ \hline
\end{tabular}
\label{tb:time}
\end{table}

\subsubsection{Discussion of Fault Cases}
\label{sc:df}

Among fault cases FT1--FT6, only FT1 is not detected by TW due to the zero fault initiating voltage angle. However, TIOC is triggered, and FT1 is still detected as a fault event. To see the performance with negative pre-fault values, cases FT6\_1, FT6\_2 and FT2\_2 are run. The algorithm shows dependability on all those fault cases, and TW elements accelerate detection compared to the overcurrent supervision element.

\subsubsection{Discussion of Switching Events}
\label{sc:ds}

Among switching events SW1--SW8, feeder energization and de-energization events SW1--SW4 and PV transformer de-energization events SW6 and SW8 do not trigger neither $TW1$ nor $TIOC$. PV transformer energization event SW5 is characterized by a large inrush current that triggers $TW1$; however, elements $TW2$ and $TI3$ are not triggered.

Cases SW2\_1--SW2\_3 and SW5\_1--SW5\_5 change parameters based on the most constraining cases, SW2 and SW5. Since SW5 is the most constraining case among all, performance against higher PV capacity is tested in SW5\_1 and SW5\_2 with variations over the parameters of SW5. Although transformer energization is usually operated at zero angle, a $10^\circ$ angle is simulated in SW5\_3 with other conditions the same as SW5, and SW5\_4 and SW5\_5 are with $90^\circ$ and $-90^\circ$ angles, respectively, to represent misoperation of switches. Performance against higher load levels is examined in SW2\_1; although it is not the normal operation, a $90^\circ$ angle is simulated in SW2\_2 for higher load levels and $-90^\circ$ in SW2\_3. Line de-energization requires low current margins; therefore, only a $0^\circ$ angle is simulated for line de-energization. The algorithm showed full security on all test switching events.

\subsubsection{Elements TW1, TW2, TI3 and TIOC in Fault Detection}
\label{sc:id}

\begin{figure}
    \centering
    \includegraphics[scale=0.2]{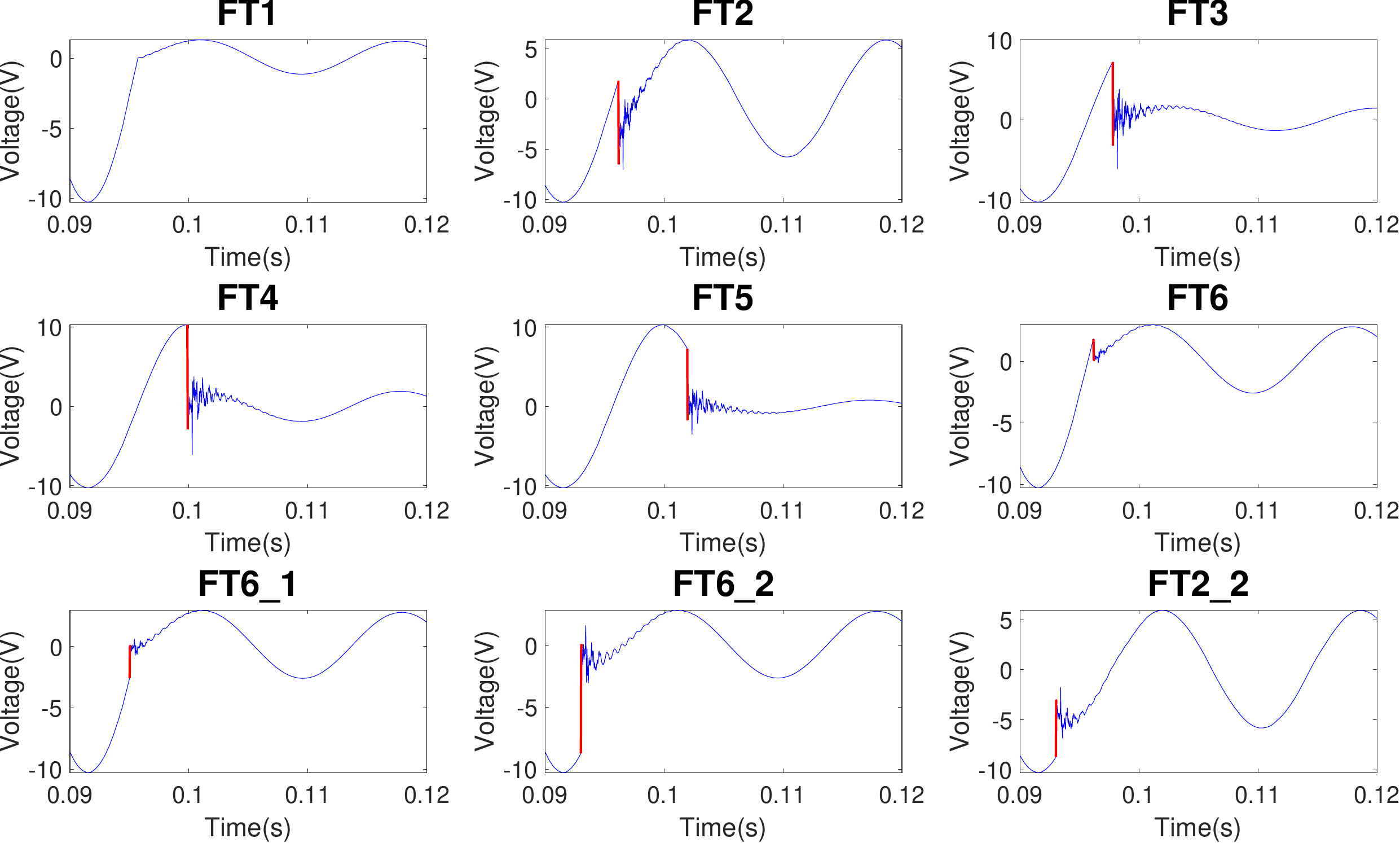}
    \caption{Voltage signals at VT1 of phase A for fault cases; red lines indicate detected TW.} 
    \label{fig:voltage_f} 
\end{figure}

Voltage signals at the voltage measurement VT1 (bus 832) of phase A for fault cases are plotted in Fig.~\ref{fig:voltage_f}; if a TW is detected in Step I, the wavefront is plotted in red.

\begin{figure}
    \centering
    \includegraphics[trim=5cm 0cm 3cm 0cm, clip,scale=0.215]{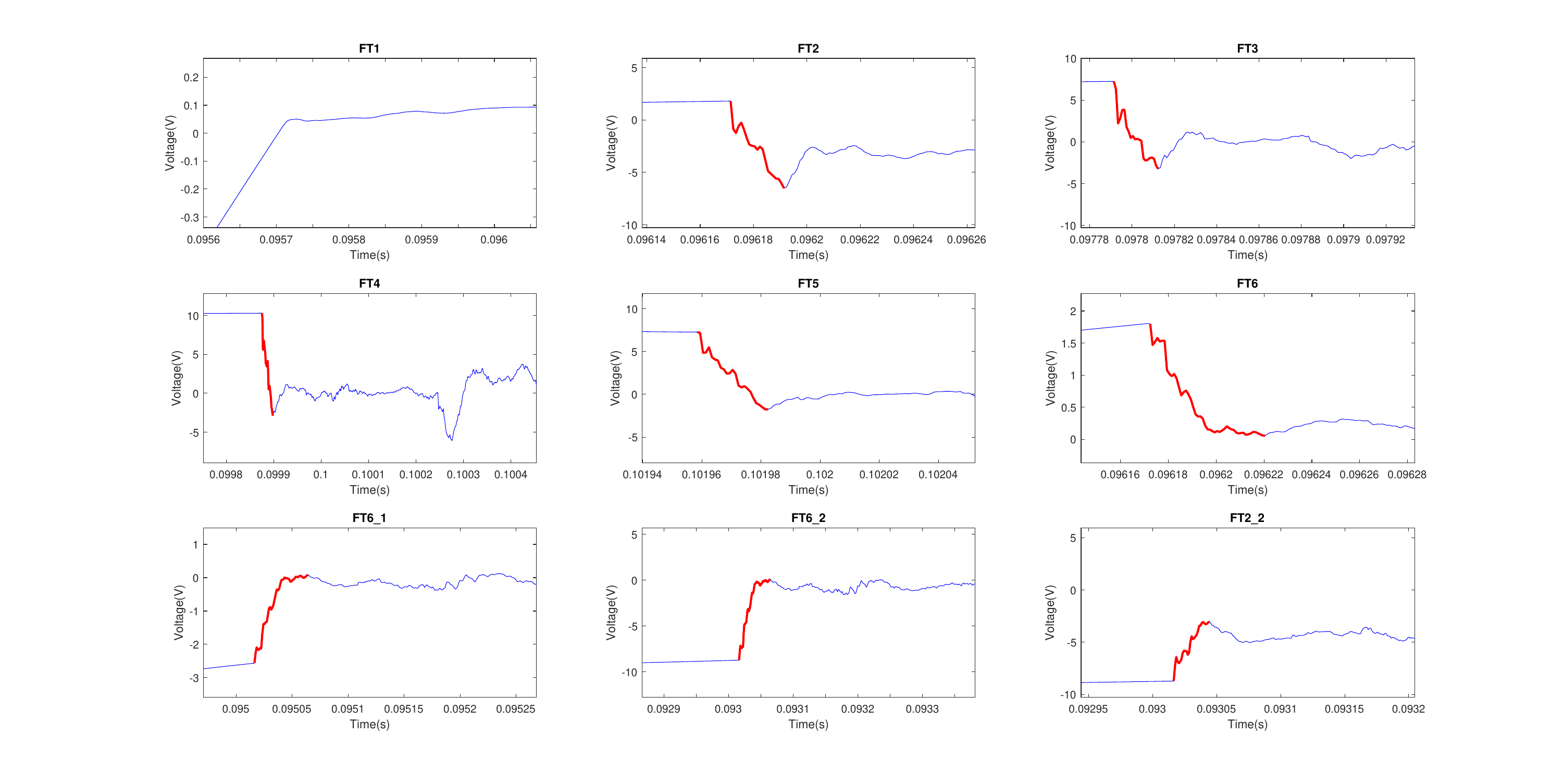}
    \caption{Detected first TW wavefronts of voltage signals at VT1 of phase A for fault cases shown in red.} 
    \label{fig:voltage_f_enlarged} 
\end{figure}

Fig. \ref{fig:voltage_f} shows that except for FT1 occurring at $0^\circ$, TWs are detected in all other fault cases, including FT2 and FT6, corresponding to a low fault initiating voltage angle. Enlarged plots of Fig.~\ref{fig:voltage_f} are given in Fig.~\ref{fig:voltage_f_enlarged} for better observation of TWs. Without the employment of filters or wavelet transformations, Step II is able to capture the wavefront of the first TWs.

\begin{figure}
    \centering
    \includegraphics[scale=0.215]{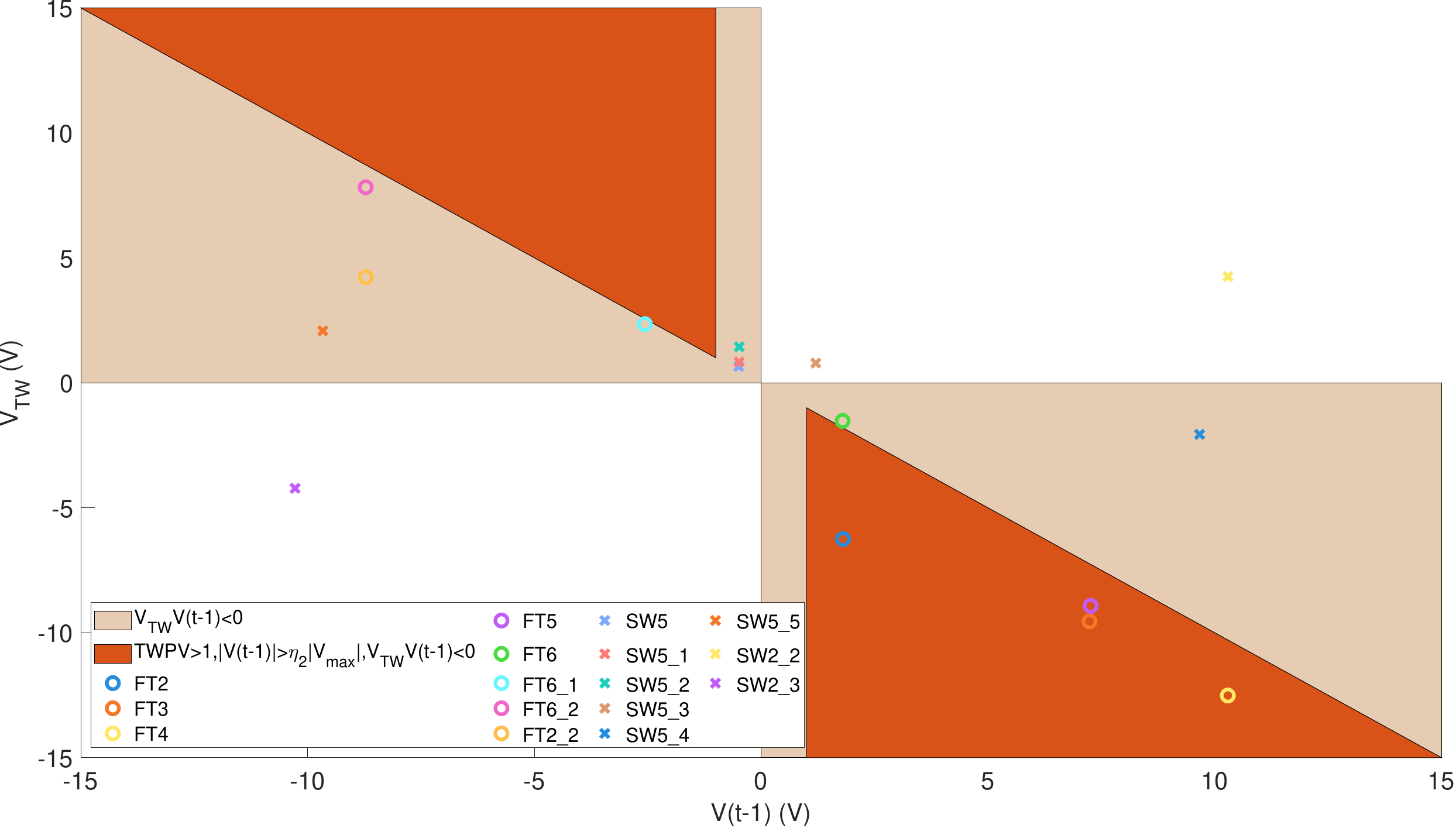}
    \caption{Pre-fault voltage values and detected TW values for fault cases and switching events; only cases triggering $TW1$ are shown.}
    \label{fig:step2}
\end{figure}

\begin{figure}
    \centering

    \subfloat[]{
        \includegraphics[width=0.45\linewidth]{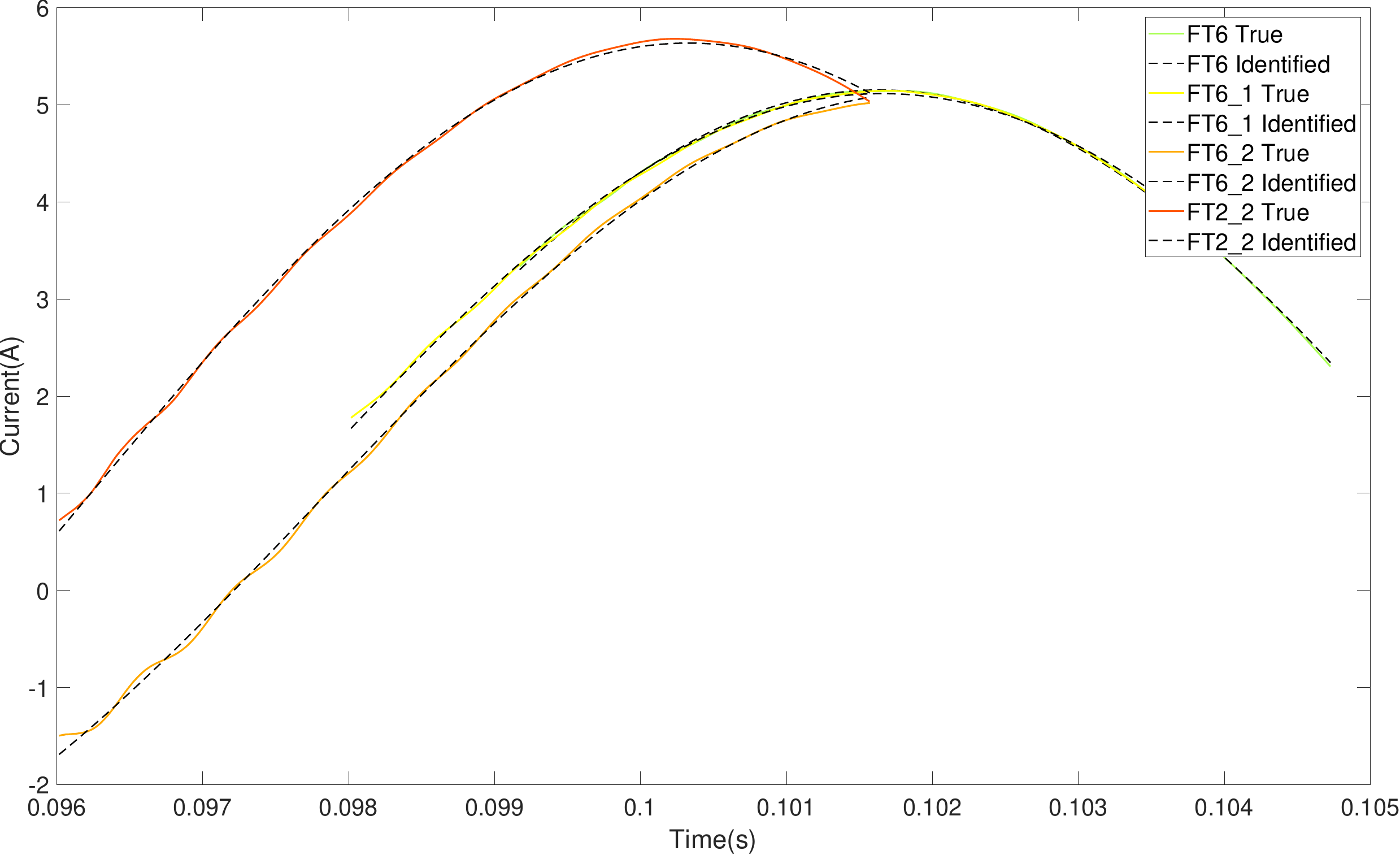}
        \label{fig:ti3_f}
    }
    \hfill
    \subfloat[]{
        \includegraphics[width=0.45\linewidth]{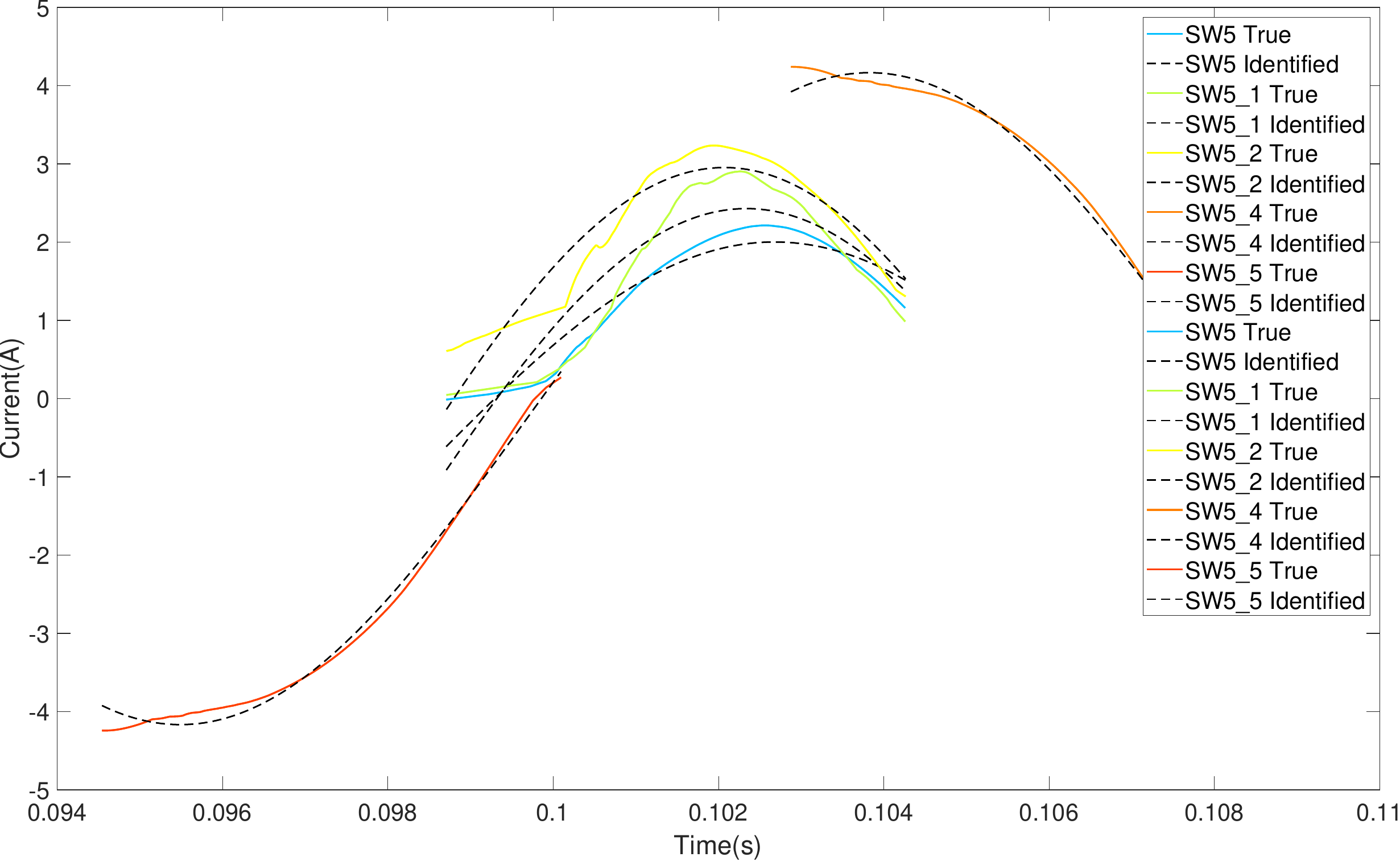}
        \label{fig:ti3_s}
    }

    \caption{Step III results for (a) fault cases (b) switching events.}
    \label{fig:ti3}
\end{figure}

\begin{figure}
    \centering

    \subfloat[]{
        \includegraphics[width=0.45\linewidth]{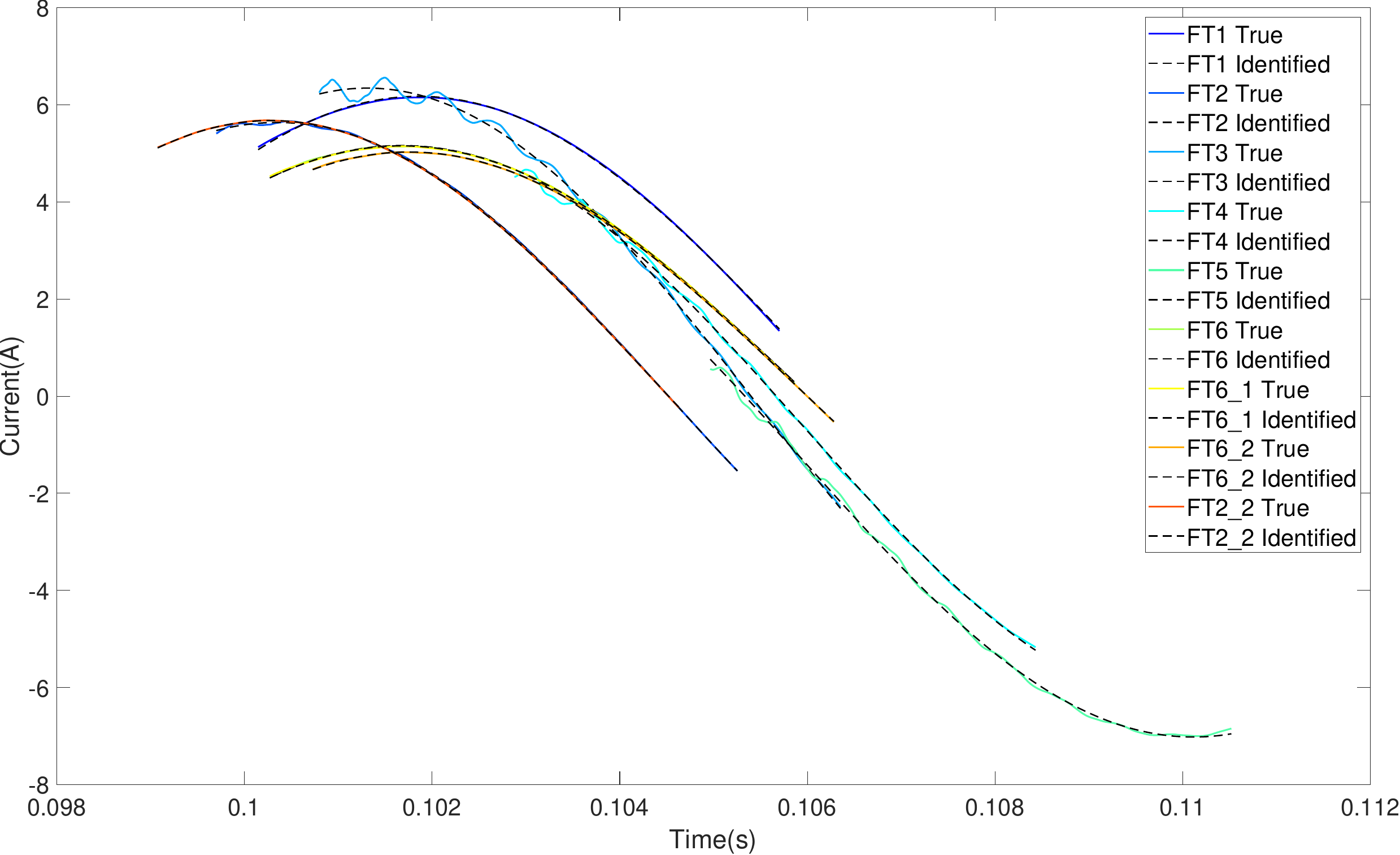}
        \label{fig:tioc_f}
    }
    \hfill
    \subfloat[]{
        \includegraphics[width=0.45\linewidth]{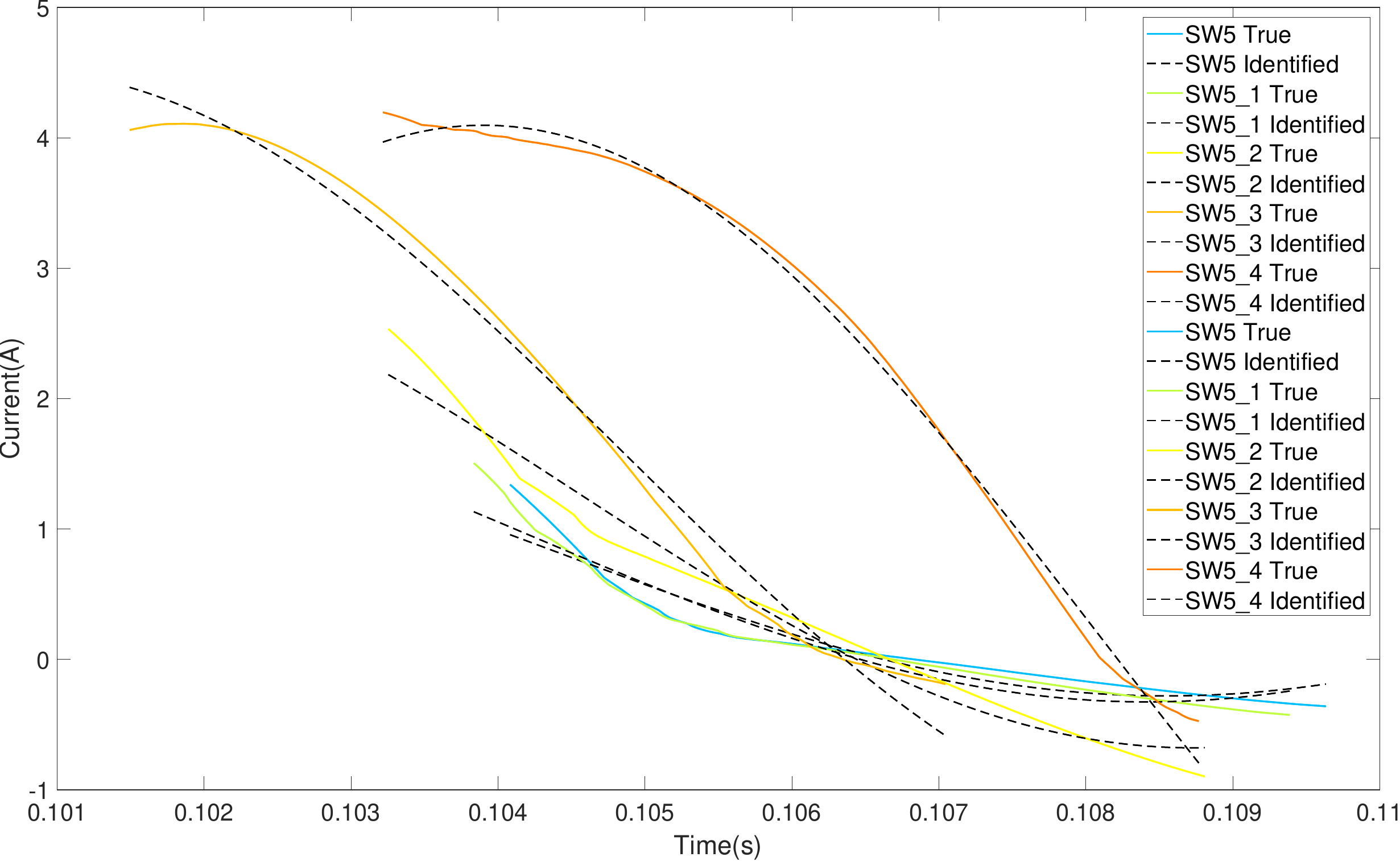}
        \label{fig:tioc_s}
    }

    \caption{Overcurrent supervision results for (a) fault cases (b) switching events.}
    \label{fig:tioc}
\end{figure}

\begin{figure}
    \centering
    \subfloat[]{
        \includegraphics[width=0.45\linewidth]{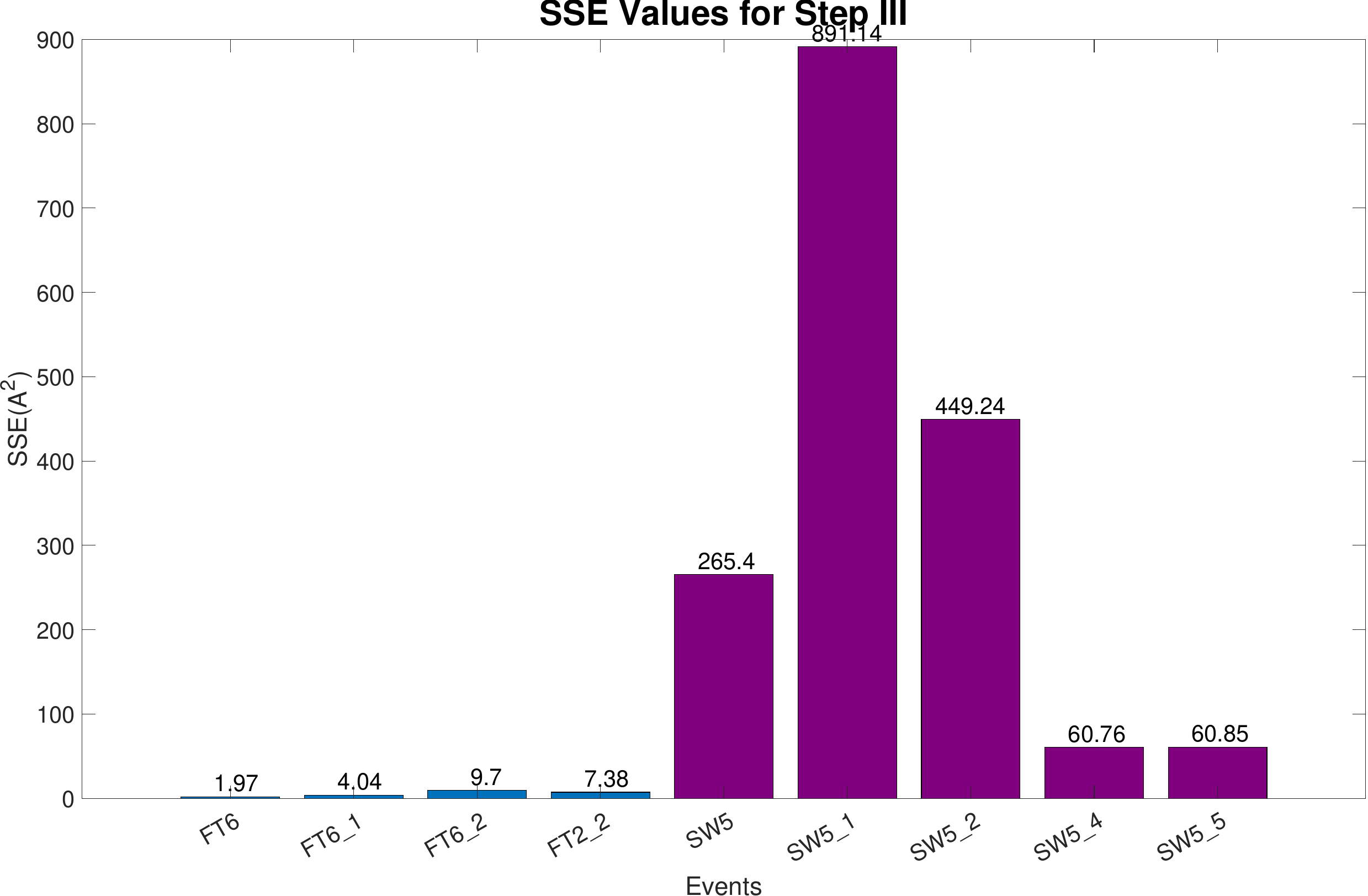}
        \label{fig:SSE_3}
    }
    \hfill
    \subfloat[]{
        \includegraphics[width=0.45\linewidth]{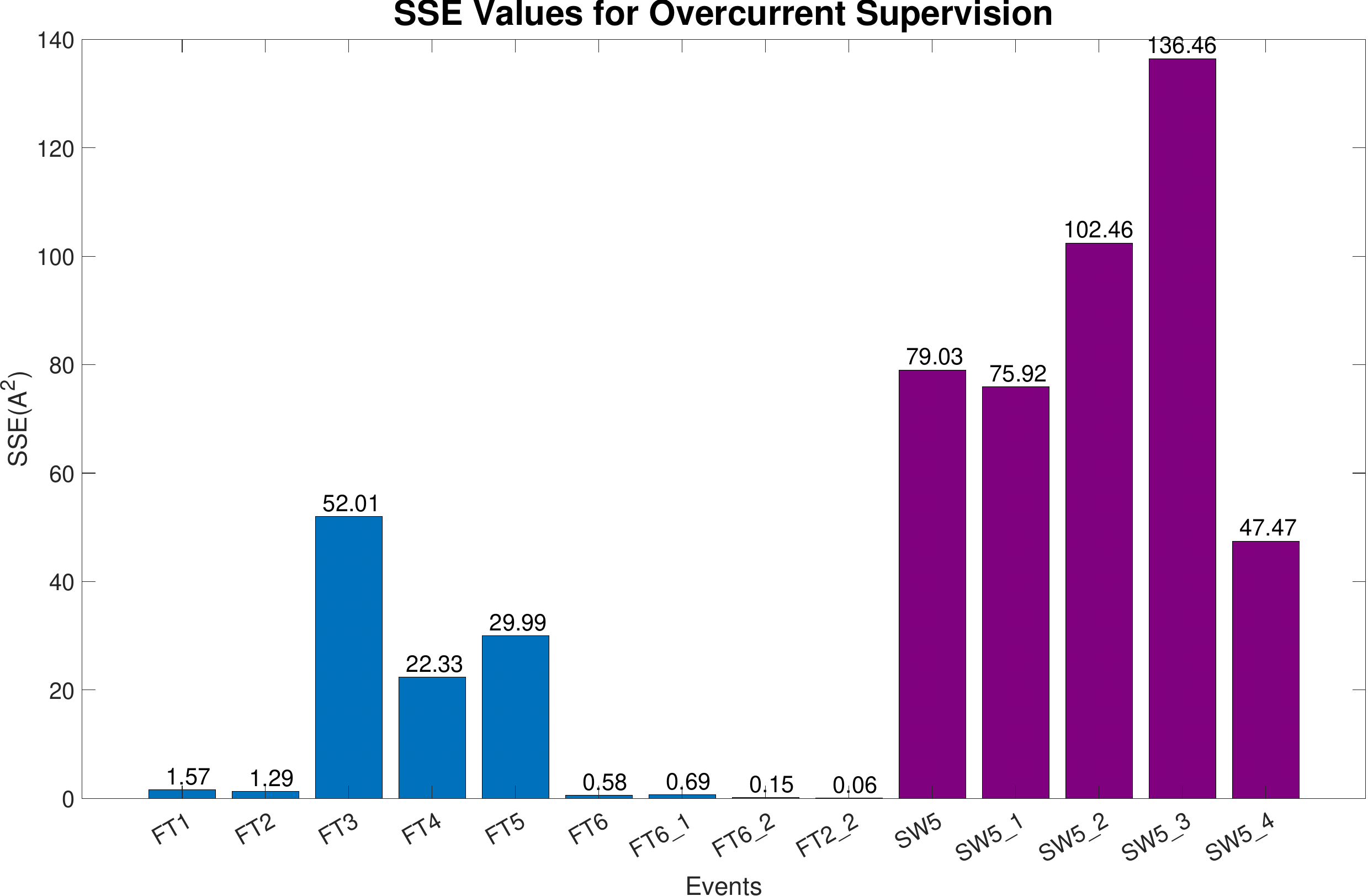}
        \label{fig:SSE_OC}
    }
    \caption{SSE values for relevant fault cases and normal switching events in (a) Step III and (b) overcurrent supervision.}
    \label{fig:SSE}
\end{figure}

For the triggered elements, Figs.~\ref{fig:step2}--\ref{fig:SSE} show details of the subsequent steps. If $TW1$ is triggered in Step I, then Step II is executed, and the TW amplitude against the pre-fault voltage value evaluated as in Fig.~\ref{fig:step2}. If $TW1$ is not triggered, then the corresponding case is not shown in Fig.~\ref{fig:step2}. Fig. \ref{fig:step2} also shows that some fault cases with higher voltage values at fault initiation thus triggering $TW2$ and $TF$. Fault cases only triggering $TW1$ are proceeded further in Step III. Fault cases not triggering $TW1$ can only rely on the overcurrent supervision for fault detection. None of the simulated switching events triggered $TW2$. Switching events triggering $TW1$, corresponding to PV transformer energization events, are proceeded further in Step III.

Fig. \ref{fig:ti3} shows the distinct behavior of PV transformer energization events compared with fault events in the time domain in Step III. Related $SSE$ values are shown in Fig.~\ref{fig:SSE}. Although detected with TWs, switching events are ruled out in Step III. Only fault events can trigger $TI3$. Fig.~\ref{fig:tioc} shows the details of overcurrent supervision corresponding to the triggering of $TIOC$. Finally, similar to Fig.~\ref{fig:ti3}, switching events are ruled out in this step. 

\section{Performance Evaluation in C-HIL}
\label{sc:HIL}

In this section, the proposed algorithm undergoes a C-HIL RTS on an islanded microgrid with an IBR. We use a lower sampling rate of 90.9~kHz (sampling time of 11 $\mu$s) imputable to hardware restrictions. Even though the sampling rate here is much lower than in commercial transmission time-domain relays \cite{SELmanual}, detection results shown next have adequate performance.

Here instead of testing over uniformly-distributed fault inception angles and checking the percentage of correct detection, we focus on low fault inception angles because the blind spots for TW methods are faults initiated around the voltage zero-crossing \cite{SELmanual, kasztenny2021traveling}. It should be noted that fault initiations at the exact voltage zero might appear possible in idealized simulations; nonetheless, in practice they cannot occur because isolation breakdown requires a minimum voltage \cite{kasztenny2023line}. In this section, we test several cases initiated with high pre-fault voltage and many cases initiated around voltage zero-crossing.

\subsection{C-HIL Implementation: Limitations and Specifications}

\begin{figure}
    \centering
    \includegraphics[scale=0.04]{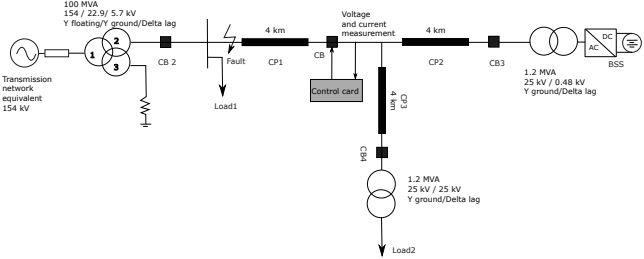}
    \caption{Microgrid for C-HIL RTS.} 
    \label{fig:rtsnet} 
\end{figure}

Simulation studies are carried out using commercial software \cite{hypersim} in its real-time mode. Although the simulator can run FPGA-based simulation, which is faster than CPU-based simulation, available system parts in the FPGA component library are limited; for example, battery storage systems (BSSs) are not available. Due to such restrictions, CPU-based RTS is done in this paper. Multiple cores are enabled to speed up the RTS. The test network shown in Fig.~\ref{fig:rtsnet} contains three line sections (CP), one BSS, three transformers, five circuit breakers (CB) and three RL loads. The BSS is run in the grid-forming (GFM) mode. Gain blocks represent the current (CT) and voltage transformers (VT) due to the limited component library, as a common practice in C-HIL setups \cite{alsharief2023power}. The gain values are $3 \times 10^{-3}$ for the current measurement and $6\times10^{-5}$ for the voltage measurement and are both added with a $1.5\,\mathrm{V}$ offset to center analog inputs to the $0\text{–}3\,\mathrm{V}$ range of the controller \cite{ticard}.

\begin{figure}
    \centering

    \subfloat[Block diagram]{%
        \includegraphics[scale=0.40,trim=3cm 2cm 3cm 8cm,clip]{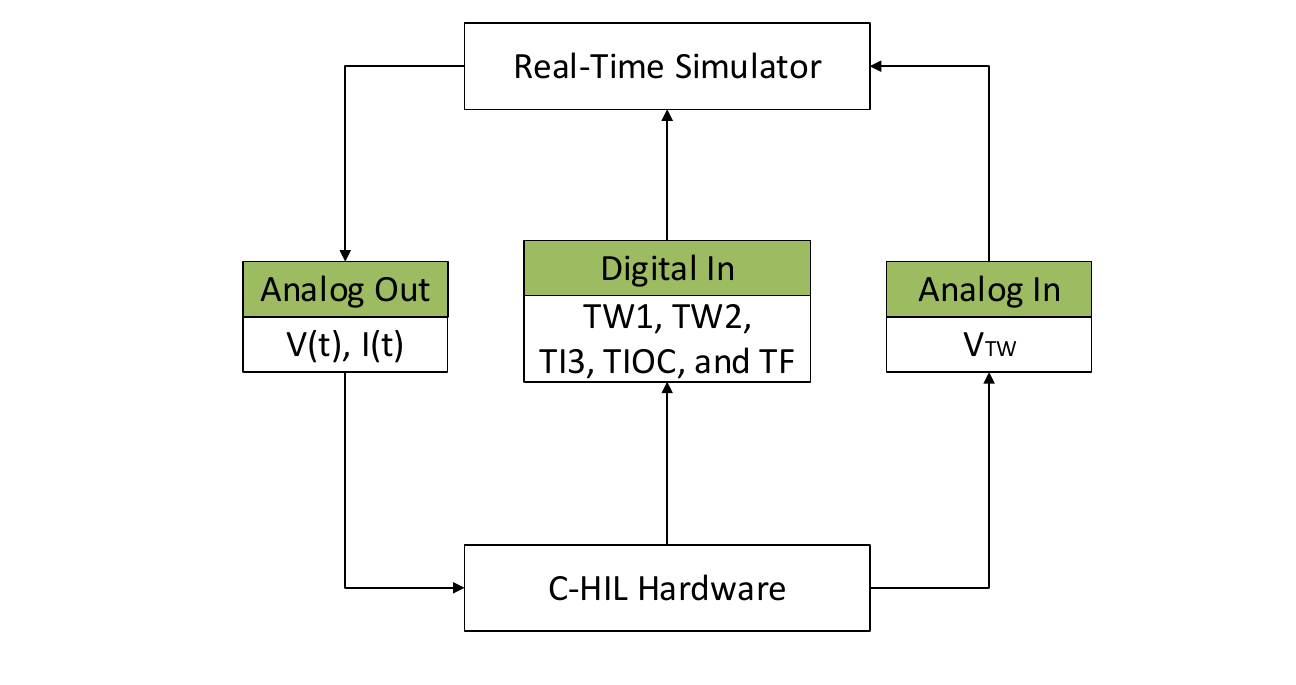}
    }

    \vspace{0.4cm}

    \subfloat[Hardware realization]{%
        \includegraphics[scale=0.30,trim=1cm 3cm 0cm 0cm,clip]{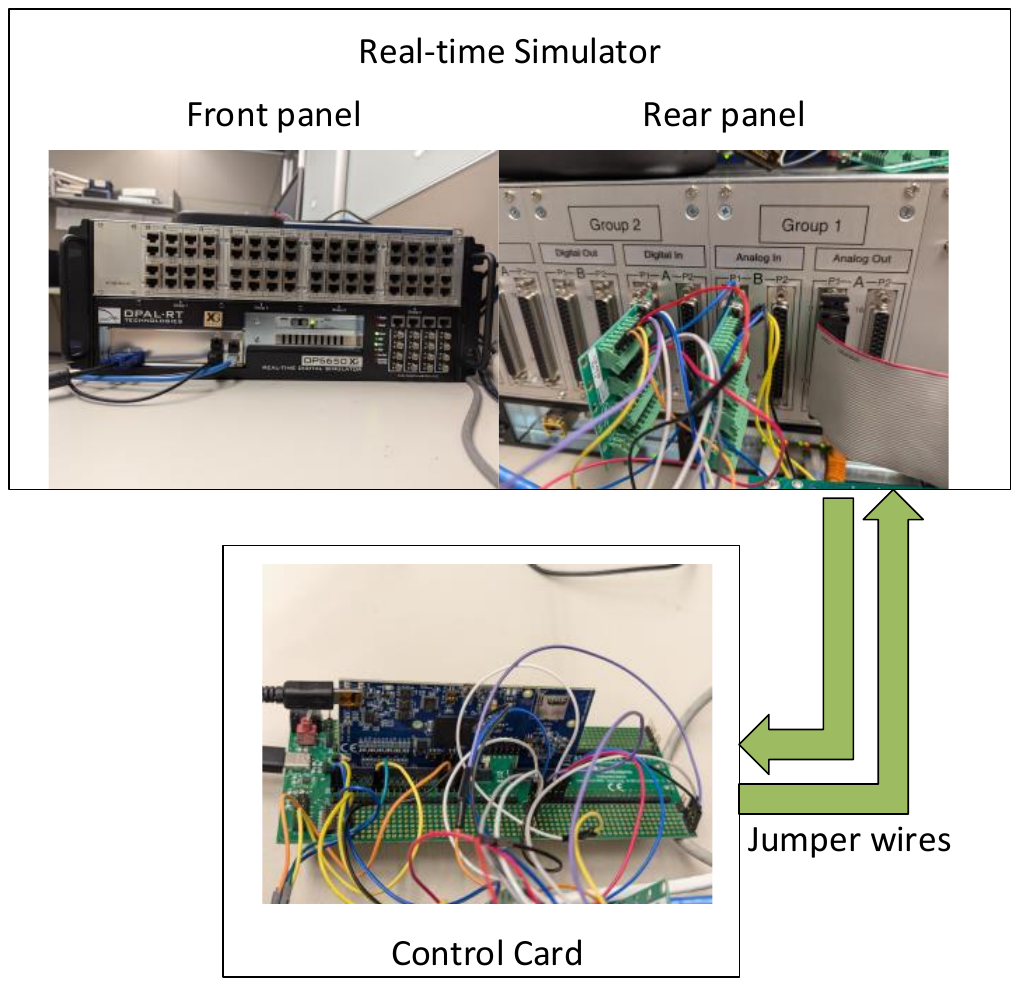}
    }

    \caption{C-HIL testing platform.}
    \label{fg:HIL}
\end{figure}

According to the execution time summary of each component, and also considering a communication delays of 4~$\mu$s across multiple cores, the minimum time step to run the network in RTS is 8.82~$\mu$s in theory. To ensure a successful RTS with the time limits, the simulation is run at a $10.5~\mu$s time step, and the control card sampling rate is set as $11 ~\mu$s. The C-HIL platform is shown as in Fig.~\ref{fg:HIL}. The simulator and the control card are connected by jumper wires.

We tested the proposed algorithm with less conservative parameter settings in comparison to those used in Section \ref{sc:V}: $\eta_1 = 1.1$, $\eta_2 = 4$, $\eta_3 = 1.5$ to emphasize the use of TW elements and rely less on backup protection.

\subsection{C-HIL RTS Results}
\label{sc:chilresults}
Table~\ref{tb:rtsft} lists a summary of fault cases with different fault types and angles. Since only phase A voltage and current are measured and investigated, all fault types related only to phase A are simulated. These results show that fault types do not affect the performance of the proposed method. Meanwhile, inverse time overcurrent relay (ANSI 51) is simulated with a US inverse curve~\cite{UScurve,8635630}, and its detection times are listed in the last column of Table~\ref{tb:rtsft}. By inspection, the ANSI 51 overcurrent protection requires much longer detection times than the proposed method.

Normal switching operations were simulated on all relevant circuit breakers to verify the security of the algorithm. At CB2, we tested if the proposed algorithm endures grid-connection of the islanded microgrid. With CB2 closed and the system running in grid-connected mode, we tested the closing of CB3. We also tested at CB4 for adding and reducing loads. None of the switching events caused $TF$ triggering and resulting CB tripping. Only the grid connection event (closing of CB2) led to observable TWs, which were not high enough to trigger breaker tripping. The algorithm was secure and dependable for all test events, and the TW elements were triggered only if faults were initiated with an angle higher than $4^\circ$. The overcurrent supervision complemented TW protection for faults with angles lower than $4^\circ$ within 10~ms, accomplishing full dependability. The next subsections look at specific cases in more detail.

\begin{table}[]
\caption{C-HIL RTS Fault Cases and Detection Times}
\begin{adjustbox}{max width=\linewidth}
\begin{tabular}{cccccccc}
\hline
\begin{tabular}[c]{@{}c@{}}Case\\ \end{tabular} & \begin{tabular}[c]{@{}c@{}}Fault \\ angle ($^\circ$)\end{tabular} & \begin{tabular}[c]{@{}c@{}}Fault \\ type\end{tabular} & \textbf{\begin{tabular}[c]{@{}c@{}}$dT_f$\\ (ms)\end{tabular}} & \begin{tabular}[c]{@{}c@{}} $dT_{TW2}$\\ (ms)\end{tabular} & \begin{tabular}[c]{@{}c@{}}$dT_{TI3}$\\ (ms)\end{tabular} & \begin{tabular}[c]{@{}c@{}} $dT_{TIOC}$\\ (ms)\end{tabular} & \begin{tabular}[c]{@{}l@{}}ANSI\\ 51 (ms)\end{tabular} \\ \hline
1                                                  & 4                                                                 & AG                                                    & \textbf{0.33}                                              & 0.33                                               & -                                                  & -                                                   & 25.69                                               \\
2                                                  & 10                                                                & AC                                                    & \textbf{0.55}                                              & 0.55                                               & -                                                  & -                                                   & \multicolumn{1}{c}{121.11}                               \\
3                                                  & 45                                                                & AG                                                    & \textbf{0.48}                                              & 0.48                                               & -                                                  & 3.80                                                & 29.81                                               \\
4                                                  & 90                                                                & AC                                                    & \textbf{0.55}                                              & 0.55                                               & -                                                  & -                                                   & 127.63                                                   \\
5                                                  & 120                                                               & ACG                                                   & \textbf{0.62}                                              & 0.62                                               & -                                                  & -                                                   & 41.19                                               \\
6                                                  & 176                                                               & AB                                                    & \textbf{0.42}                                              & 0.42                                               & -                                                  & -                                                   & 124.03                                                   \\
7                                                  & 180                                                               & AG                                                    & \textbf{8.40}                                              & -                                                  & -                                                  & 8.40                                                & 47.36                                               \\
8                                                  & 184                                                               & AG                                                    & \textbf{9.00}                                              & -                                                  & -                                                  & 9.00                                                & 48.36                                               \\
9                                                  & 240                                                               & ABCG                                                  & \textbf{0.49}                                              & 0.49                                               & -                                                  & -                                                   & 88.23                                               \\
10                                                 & 270                                                               & ABG                                                   & \textbf{0.65}                                              & 0.65                                               & -                                                  & -                                                   & 45.30                                               \\
11                                                 & 315                                                               & ABC                                                   & \textbf{0.40}                                              & 0.40                                               & -                                                  & -                                                   & 86.23                                               \\
12                                                 & 350                                                               & AB                                                    & \textbf{0.95}                                              & 0.95                                               & -                                                  & -                                                   & 121.25                                                   \\
13                                                 & 356                                                               & AG                                                    & \textbf{0.81}                                              & 0.81                                               & -                                                  & -                                                   & 26.06                                               \\
14                                                 & 357                                                               & AG                                                    & \textbf{9.00}                                              & -                                                  & -                                                  & 9.00                                                & 26.01                                               \\ \hline

\end{tabular}
\end{adjustbox}
\label{tb:rtsft}
\end{table}

\subsubsection{Case~3: $45^\circ$ AG fault in Islanded Operation}
\label{sc:rtft3}

\begin{figure}
    \centering
    \includegraphics[scale=0.32]{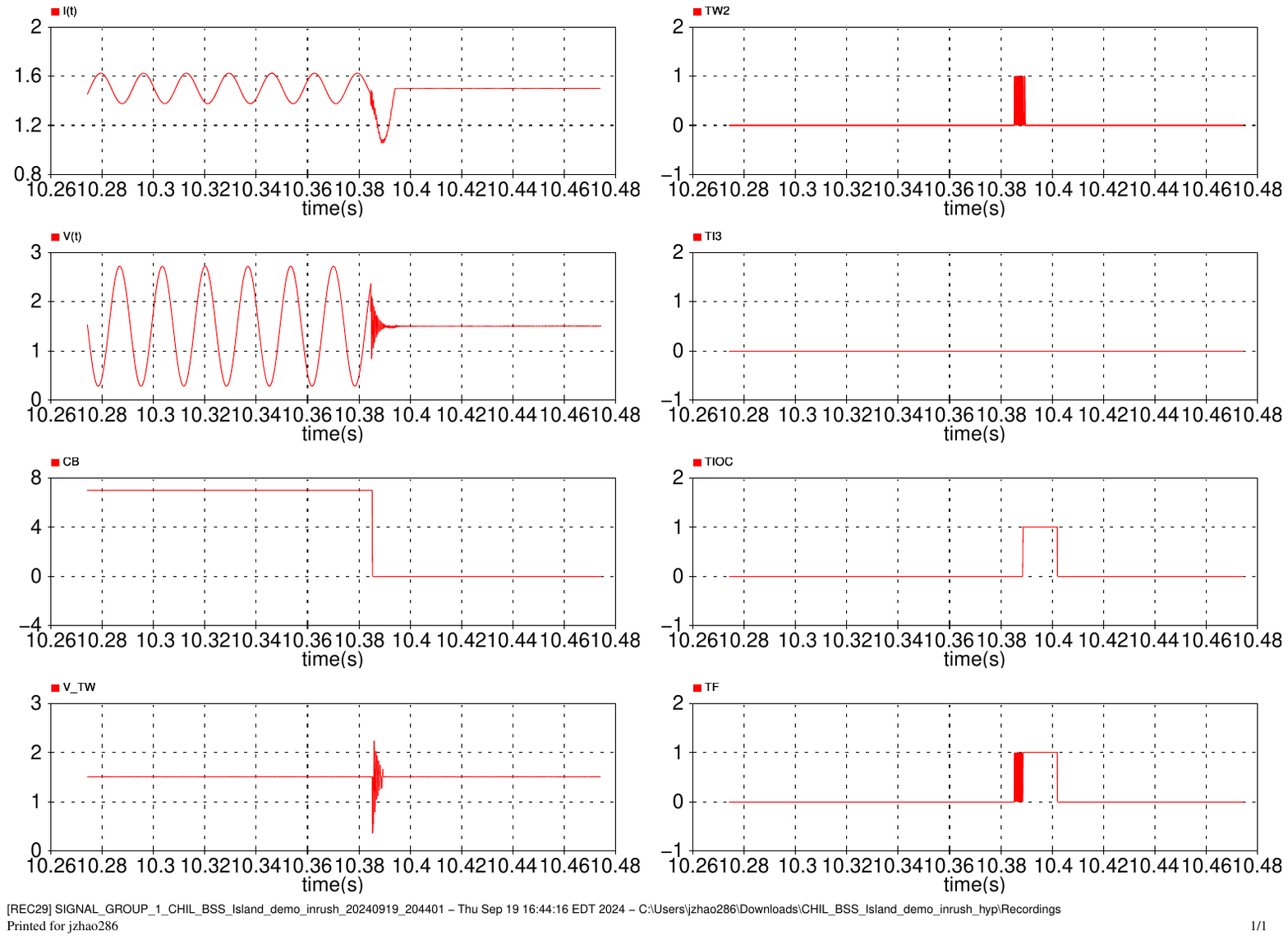}
    \caption{C-HIL results for an AG fault with a $45^\circ$ inception angle.}
    \label{fig:rts45}
\end{figure}

\begin{figure}
    \centering
    \includegraphics[scale=0.32]{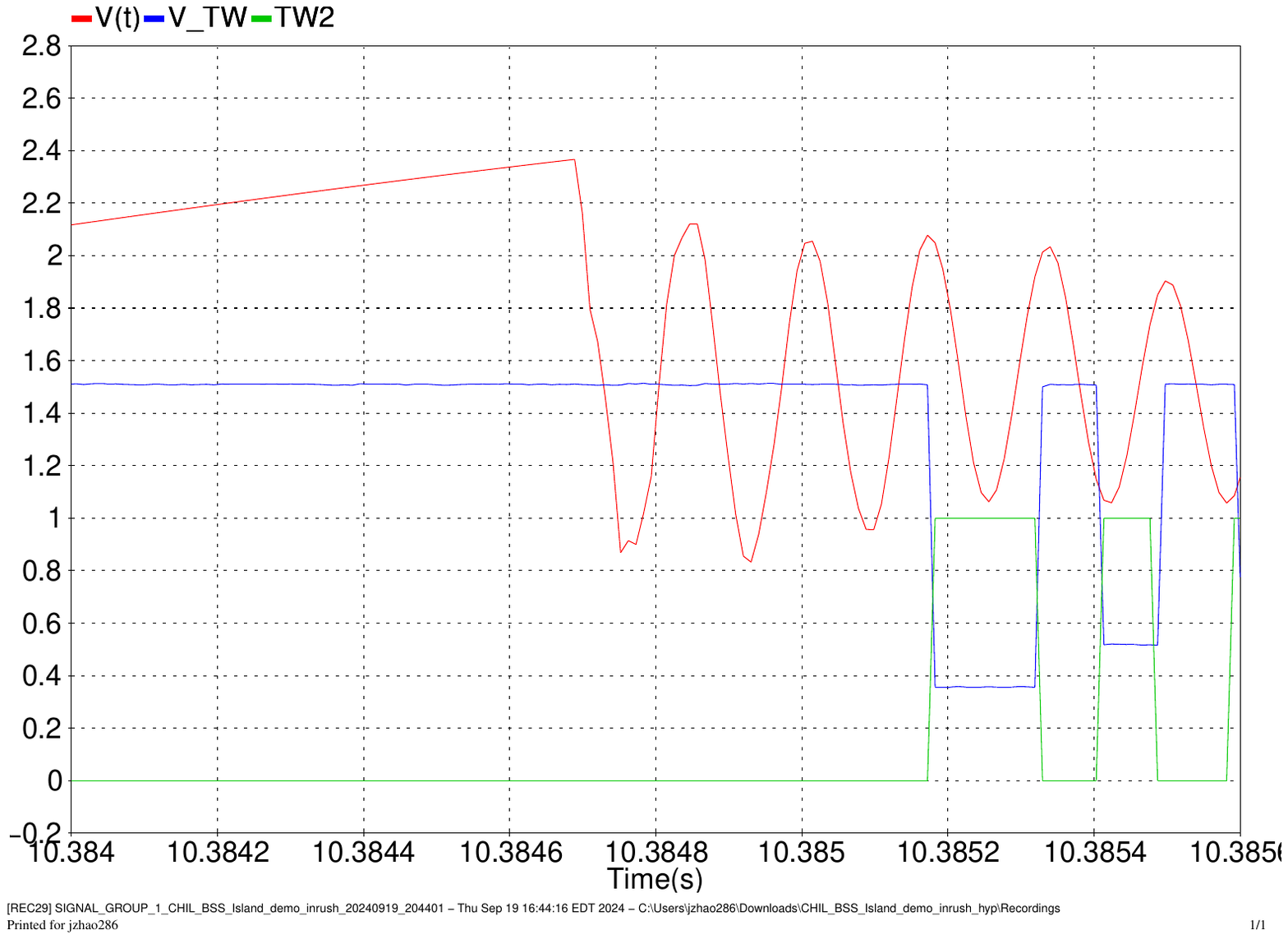}
    \caption{Zoom in of $V(t)$, $V_{TW}$ and $TW2$ around the 45$^\circ$ AG fault inception.}
    \label{fig:enlargeTW}
\end{figure}

Here phase A was faulted to ground with an inception angle of $45^\circ$ at the fault location in Fig.~\ref{fig:rtsnet}; Fig.~\ref{fig:rts45} provides corresponding simulator inputs and outputs. Fig.~\ref{fig:enlargeTW} is an enlarged plot of $V_{TW}$, $V(t)$ and $TW2$. When the fault initiated at 10.38469~s, $TW2$ and $TF$ changed from 0 to 1 at 10.38517~s, causing the circuit breaker state to change from three-phase closed (value 7) to open (value 0). Signals $V(t)$ and $I(t)$ in Fig.~\ref{fig:rts45} are output signals of the simulator. Due to the existence of noise in the controller, the measured TW amplitude $V_{TW}$ is slightly lower than its true value. Also, $TIOC$ changes to 1 at 10.38851~s.

\subsubsection{Case 4: $4^\circ$ AG fault in Islanded Operation}
\label{sc:rtft4}

\begin{figure}
    \centering
    \includegraphics[scale=0.32]{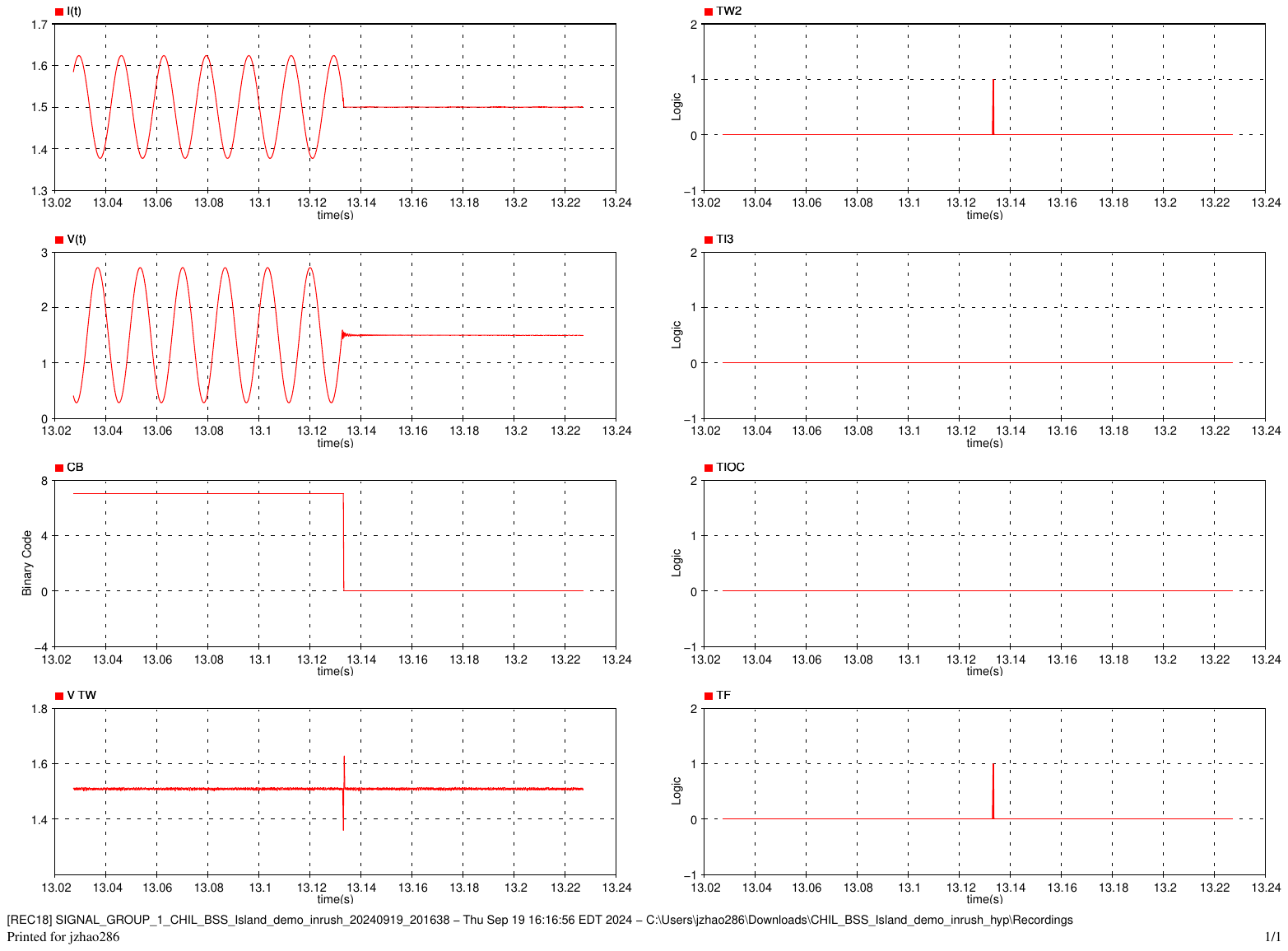}
    \caption{CC-HIL results for an AG fault with a $4^\circ$ inception angle.}
    \label{fig:rts4}
\end{figure}

The same fault as above was applied, but with a $4^\circ$ inception angle. Fig.~\ref{fig:rts4} shows the simulator outputs and inputs. When the fault was initiated at 13.13280~s, $TW2$ and $TF$ changed from 0 to 1 at 13.13313~s, causing the circuit breaker state to change from three-phase closed to open. Because of the low fault current magnitude here, $TIOC$ was never triggered.

\subsubsection{Normal Switching of CB2 from Islanded to Grid Connected}
\label{sc:cb2}
\begin{figure}
    \centering
    \includegraphics[scale=0.32]{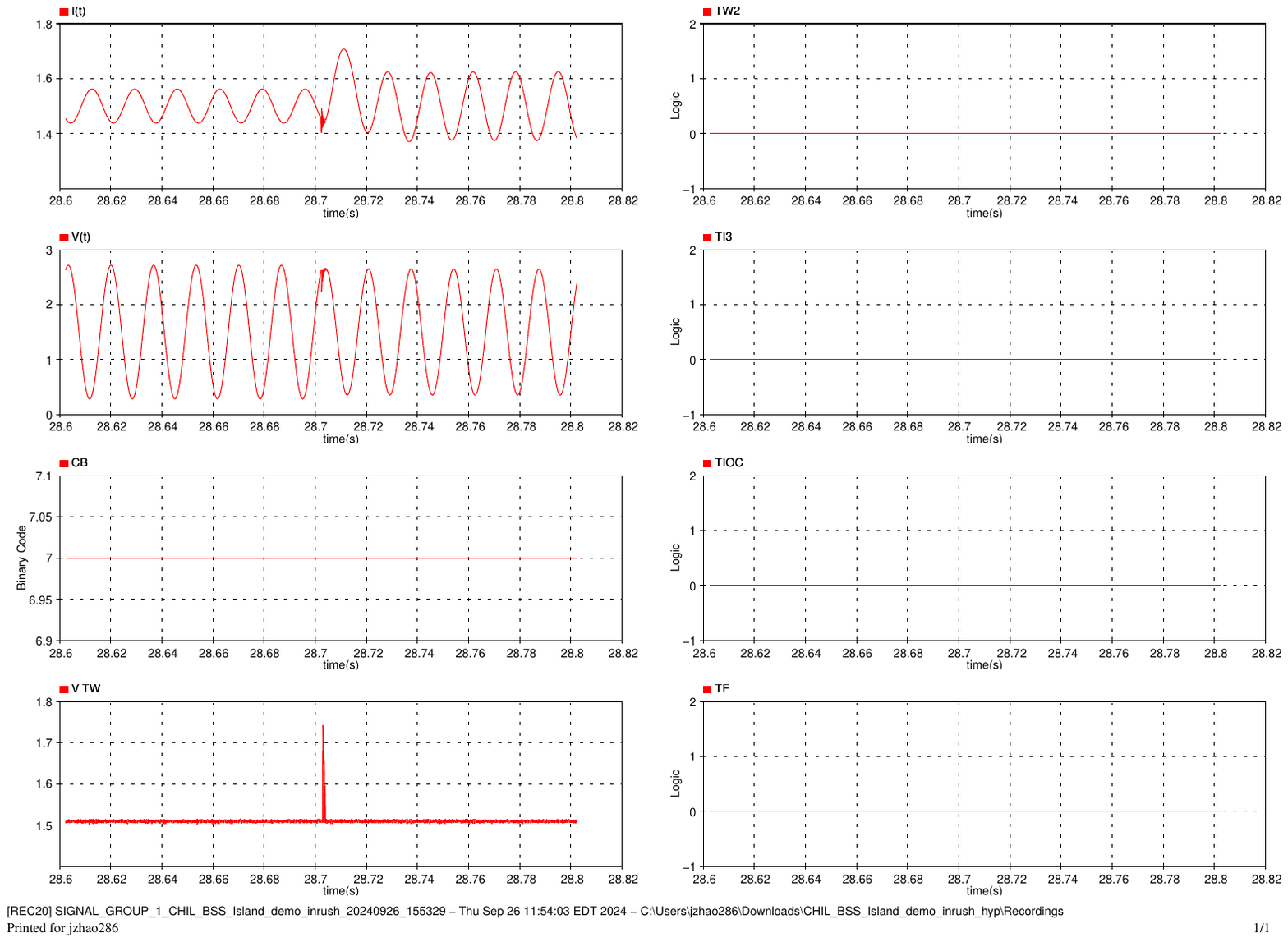}
    \caption{C-HIL results for a normal switching event at CB2.}
    \label{fig:CB2}
\end{figure}

As shown in Fig.~\ref{fig:CB2}, a TW signal is detected at 28.70290~s after CB2 is closed at 28.70239~s. However, this TW amplitude was not high enough compared with the pre-fault voltage, leading to $TW2$ not being triggered and leaving $TF = 0$. 

\section{Conclusion}\label{conclusion}
\label{sc:conclusion}

This paper proposed a pre-fault voltage discrimination strategy and a new time-domain fault detection algorithm for distribution networks with IBRs. 
The algorithm utilizes TW information for fast fault detection while not making a compromise between dependability and security. The proposed algorithm was tested with both offline simulation and C-HIL RTS. Of major importance for fault detection, the dependability of the algorithm against small fault angles and its security against normal switching cases were validated. With additional merits, its protection speed is higher than traditional overcurrent relays due to the use of traveling waves and other time-domain information. Future work is addressing the extension of the method to multi-ended or communication assisted protection schemes, as well as its application in meshed or reconfigurable networks. Additional studies will examine interactions with diverse inverter control schemes and assess robustness under practical measurement limitations.

\bibliographystyle{IEEEtran}
\typeout{}
\bibliography{references}

@inproceedings{schweitzer2015speed,
  title={Speed of line protection-can we break free of phasor limitations?},
  author={Schweitzer, Edmund O and Kasztenny, Bogdan and Guzm{\'a}n, Armando and Skendzic, Veselin and Mynam, Mangapathirao V},
  booktitle={2015 68th Annual Conference for Protective Relay Engineers},
  pages={448--461},
  year={2015},
  organization={IEEE}
}

@article{de2021mathematical,
  title={Mathematical study on traveling waves phenomena on three phase transmission lines--{Part I}: Fault-Launched Waves},
  author={de Magalh{\~a}es J{\'u}nior, Fernando Marinho and Lopes, Felipe V},
  journal={IEEE Transactions on Power Delivery},
  volume={37},
  number={2},
  pages={1151--1160},
  year={2021},
  publisher={IEEE}
}

@book{greenwood1991electrical,
  title={Electrical Transients in Power Systems},
  author={Greenwood, Allan},
  year={1991},
  publisher={New York, N.Y.: John Wiley and Sons}
}

@book{kimbark1949electrical,
  title={Electrical transmission of power and signals},
  author={Kimbark, Edward Wilson},
  publisher={London, UK: John Wiley},
  year={1949}
}

@article{michalik2006high,
  title={High-impedance fault detection in distribution networks with use of wavelet-based algorithm},
  author={Michalik, Marek and Rebizant, Waldemar and Lukowicz, Miroslaw and Lee, Seung-Jae and Kang, Sang-Hee},
  journal={IEEE Transactions on Power Delivery},
  volume={21},
  number={4},
  pages={1793--1802},
  year={2006},
  publisher={IEEE}
}

@article{sarwagya2018high,
  title={High-impedance fault detection in electrical power distribution systems using moving sum approach},
  author={Sarwagya, Kumari and De, Sourav and Nayak, Paresh Kumar},
  journal={IET Science, Measurement \& Technology},
  volume={12},
  number={1},
  pages={1--8},
  year={2018},
  publisher={Wiley Online Library}
}

@article{lin2019discrete,
  title={Discrete wavelet transform-based triggering method for single-phase earth fault in power distribution systems},
  author={Lin, Cheng and Gao, Wei and Guo, Mou-Fa},
  journal={IEEE Transactions on Power Delivery},
  volume={34},
  number={5},
  pages={2058--2068},
  year={2019},
  publisher={IEEE}
}

@techreport{prabakar2021use,
  title={Use of traveling wave signatures in medium-voltage distribution systems for fault detection and location},
  author={Prabakar, Kumaraguru and Singh, Akanksha and Reynolds, Matthew and Lunacek, Monte and Monz{\'o}n, Lucas and Velaga, Yaswanth Nag and Maack, Jonathan and Tiwari, Soumya and Roy, Jinia and Tombari, Colin and others},
  year={2021},
  institution={National Renewable Energy Laboratory, Golden, CO}
}

@inproceedings{schweitzer2016new,
  title={New time-domain line protection principles and implementation},
  author={Schweitzer, E.O. and Kasztenny, B and Mynam, M and Guzman, A and Skendzic, V},
  booktitle={Proc. 13th International Conference on Developments in Power System Protection},
  pages={1--6},
  year={2016}
}

@misc{ieee34, 
  author = {{IEEE PES AMPS DSAS Test Feeder Working Group}}, 
  title = {{IEEE 34} node test feeder}, 
  howpublished = {\url{http://sites.ieee.org/pes-testfeeders/resources/}}, 
  lastchecked = {05.09.2023}, 
  originalyear = {2008} 
}

@book{van2001transients,
  title={Transients in power systems},
  author={Van der Sluis, Lou},
  year={2001},
  publisher={New York, NY: Wiley}
}

@misc{SELmanual, 
  author = {{Schweitzer Engineering Laboratories}}, 
  title = {Instruction Manual of {SEL-T400L} Time Domain Line Protection}, 
  howpublished = {\url{https://selinc.com/products/T400L/docs/}}, 
  lastchecked = {05.09.2023}, 
  originalyear = {2017}
}

@inproceedings{schweitzer2016performance,
  title={Performance of time-domain line protection elements on real-world faults},
  author={Schweitzer, Edmund O and Kasztenny, Bogdan and Mynam, Mangapathirao V},
  booktitle={Proc. 69th Annual Conference for Protective Relay Engineers},
  pages={1--17},
  year={2016}
}

@article{galvez2020fault,
  title={Fault location in active distribution networks containing distributed energy resources ({DERs})},
  author={Galvez, Cesar and Abur, Ali},
  journal={IEEE Transactions on Power Delivery},
  volume={36},
  number={5},
  pages={3128--3139},
  year={2020},
  publisher={IEEE}
}

@inproceedings{schweitzer2014locating,
  title={Locating faults by the traveling waves they launch},
  author={Schweitzer, Edmund O and Guzm{\'a}n, Armando and Mynam, Mangapathirao V and Skendzic, Veselin and Kasztenny, Bogdan and Marx, Stephen},
  booktitle={Proc. 67th Annual Conference for Protective Relay Engineers},
  pages={95--110},
  year={2014}
}

@inproceedings{mansourlakouraj2021application,
  title={Application of Graph Neural Network for Fault Location in {PV} Penetrated Distribution Grids},
  author={MansourLakouraj, Mohammad and Hossain, Rakib and Livani, Hanif and Ben-Idris, Mohammed},
  booktitle={2021 North American Power Symposium},
  pages={1--6},
  year={2021}
}

@article{vaish2021machine,
  title={Machine learning applications in power system fault diagnosis: Research advancements and perspectives},
  author={Vaish, Rachna and Dwivedi, UD and Tewari, Saurabh and Tripathi, SM},
  journal={Engineering Applications of Artificial Intelligence},
  volume={106},
  pages={104504},
  year={2021},
  publisher={Elsevier}
}

@article{tashakkori2019fault, 
  title={Fault location on radial distribution networks via distributed synchronized traveling wave detectors}, 
  author={Tashakkori, Ali and Wolfs, Peter J and Islam, Syed and Abu-Siada, Ahmed}, 
  journal={IEEE Transactions on Power Delivery}, 
  volume={35}, 
  number={3}, 
  pages={1553--1562}, 
  year={2019}, 
  publisher={IEEE} 
}

@inproceedings{dommel1978high,
    author = {Dommel, Hermann W.},
    title = {High speed relaying using traveling wave transient analysis},
    booktitle ={Proc. IEEE PES Winter Meeting},
    year = {1978}
}

@article{shehab1988travelling,
  title={Travelling wave distance protection-problem areas and solutions},
  author={Shehab-Eldin, EH and McLaren, PG},
  journal={IEEE Transactions on Power Delivery},
  volume={3},
  number={3},
  pages={894--902},
  year={1988},
  publisher={IEEE}
}

@article{crossley1983distance,
  title={Distance protection based on travelling waves},
  author={Crossley, PA and McLaren, PG},
  journal={IEEE Transactions on Power Apparatus and Systems},
  volume = {PAS-102},
  number={9},
  pages={2971--2983},
  year={1983},
  publisher={IEEE}
}

@article{magnago1998fault,
  title={Fault location using wavelets},
  author={Magnago, Fernando H and Abur, Ali},
  journal={IEEE Transactions on Power Delivery},
  volume={13},
  number={4},
  pages={1475--1480},
  year={1998},
  publisher={IEEE}
}

@article{vitins1981fundamental, 
  title={A fundamental concept for high speed relaying}, 
  author={Vitins, M}, 
  journal={IEEE Transactions on Power Apparatus and Systems},
  volume={PAS-100},
  number={1}, 
  pages={163--173}, 
  year={1981}, 
  publisher={IEEE} 
}

@article{borghetti2008continuous,
  title={Continuous-wavelet transform for fault location in distribution power networks: Definition of mother wavelets inferred from fault originated transients},
  author={Borghetti, Alberto and Bosetti, Mauro and Di Silvestro, Mauro and Nucci, Carlo Alberto and Paolone, Mario},
  journal={IEEE Transactions on Power Systems},
  volume={23},
  number={2},
  pages={380--388},
  year={2008},
  publisher={IEEE}
}

@inproceedings{kasztenny2021traveling,
  title={Traveling-Wave Overcurrent--A New Way to Protect Lines Terminated on Transformers},
  author={Kasztenny, Bogdan and Mynam, Mangapathirao V and Marx, Stephen and Barone, Ralph},
  booktitle={Proc. 48th Annual Western Protective Relay Conference},
  year={2021}
}

@techreport{muenz2024protection,
  title={Protection of 100\% inverter-dominated power systems with grid-forming inverters and protection relays--gap analysis and expert interviews},
  author={Muenz, Ulrich and Bhela, Siddharth and Xue, Nan and Banerjee, Abhishek and Reno, Matthew J and Kelly, Daniel James and Farantatos, Evangelos and Haddadi, Aboutaleb and Ramasubramanian, Deepak and Banaie, Amin},
  year={2024},
  institution={Sandia National Laboratory, Albuquerque, NM}
}

@article{kasztenny2023line,
  title={Line Protective Relays Suitable for Systems With a High Penetration of Unconventional Sources--Operating Principles and Field Experience},
  author={Kasztenny, B},
  journal={Cigre Science \& Engineering},
  number={31},
  pages={1--14},
  year={2023}
}

@inproceedings{alsharief2023power,
  title={Power System Modeling for Hardware-In-the-Loop Testing of A Traveling-Wave-Based Relay},
  author={Alsharief, Yagoob and Ijaz, Muhammed and Haddadi, Aboutaleb and Xue, Mark and Kadavil, Rahul},
  booktitle={Proc. IEEE PES Innovative Smart Grid Technologies Latin America},
  pages={465--469},
  year={2023},
}

@misc{emtprv, 
  author = {Jean Mahseredjian and Chris Dewhurst}, 
  title = {{EMTP} User Manual Version 4.4}, 
  howpublished = {\url{https://emtp.com/documentation/emtp-user-manual}}, 
  year = {2024} 
}

@misc{hypersim, 
  author = {{OPAL-RT Technologies, Inc.}}, 
  title = {{HYPERSIM} Documentation}, 
  howpublished = {\url{https://opal-rt.atlassian.net/wiki/spaces/PDOCHS/overview?homepageId=150372817}}, 
  year = {2025}
}

@misc{ticard, 
  author = {{Texas Instruments}}, 
  title = {User’s Guide {Delfino TMS320F28379D controlCARD R1.3}}, 
  howpublished = {\url{https://www.ti.com/lit/ug/sprui76b/sprui76b.pdf?ts=1732344677331}}, 
  year = {2022}
}

@inproceedings{kasztenny2023dependability,
author = {Kasztenny, Bogdan},
booktitle={Proc. Protection, Automation \& Control World Conf.},
year = {2023},
pages = {1--12},
title = {Dependability of Transient-Based Line Protection Elements and Schemes}
}

@misc{UScurve, 
  author = {{OPAL-RT Technologies, Inc.}},
  title = {{50-51 - Overcurrent Relay}}, 
  howpublished = {\url{https://opal-rt.atlassian.net/wiki/spaces/PDOCHS/pages/150244978/50-51+-+Overcurrent+Relay/}}, 
  year = {2023} 
}

@ARTICLE{8635630,
  author={},
  journal={{IEEE} Std. C37.112-2018}, 
  title={{IEEE} Standard for Inverse-Time Characteristics Equations for Overcurrent Relays}, 
  year={2019},
  volume={},
  number={},
  pages={1-25},
  doi={10.1109/IEEESTD.2019.8635630}}

\end{document}